\documentclass[11pt]{article}
\usepackage{amsmath,epsfig,graphicx,amssymb,bm}
\begin{document}

\thispagestyle{empty}

\begin{center}

{\Huge\bf Near-field Electrodynamics of\\
Atomically Doped Carbon\\[0.25cm] Nanotubes$^{\ast}$}

\vskip1cm

{\Large\bf Igor V. Bondarev$^{\dag,1}$, ~Philippe Lambin$^\ddag$}

\vskip1cm

{\large {\it $^\dag$ Institute for Nuclear Problems,\\ Belarusian
State University,\\ Bobruiskaya Str.11, 220050 Minsk, BELARUS}} \\[0.5cm]

{\large {\it $^\ddag$ Facult\'{e}s Universitaires Notre-Dame de la
Paix,\\ 61 rue de Bruxelles, 5000 Namur, BELGIUM}}

\end{center}

\vskip1cm

\begin{minipage}{5in}
\centerline{{\sc Abstract}}
\medskip

We develop a quantum theory of near-field electrodynamical
properties of carbon nanotubes and investigate spontaneous decay
dynamics of excited states and van der Waals attraction of the
ground state of an atomic system close to a single-wall nanotube
surface. Atomic spontaneous decay exhibits vacuum-field Rabi
oscillations -- a principal signature of strong atom-vacuum-field
coupling. The strongly coupled atomic state is nothing but
a~'quasi-1D cavity polariton'. Its stability is mainly determined
by the atom-nanotube van der Waals interaction. Our calculations
of the ground-state atom van der Waals energy performed within a
universal quantum mechanical approach valid for both weak and
strong atom-field coupling demonstrate the inapplicability of
conventional weak-coupling-based van der Waals interaction models
in a close vicinity of the nanotube surface.
\end{minipage}

\vskip 0.5cm


\vskip 1.7cm

\noindent$\overline{~~~~~~~~~~~~~~~~~~~~~~~~~~~~~~~~~}$

\noindent$^{\ast}$~To be published in \emph{"Nanotubes: New
Research"}, edited by F.Columbus (Nova $\mbox{~~}$~Science, New
York, 2005).

\noindent$^{1}$~Corresponding author. E-mail address: {\tt
bondarev@tut.by}

\newpage

\tableofcontents

\newpage

\section{Introduction}

Carbon nanotubes (CNs) are graphene sheets rolled-up into
cylinders of approximately one nanometer in diameter.~Extensive
work carried out worldwide in recent years has revealed the
intriguing physical properties of these novel molecular scale
wires~\cite{Dresselhaus,Dai}.~Nanotubes have been shown to be
useful for miniaturized electronic, mechanical, electromechanical,
chemical and scanning probe devices and materials for macroscopic
composites~\cite{Baughman}.~Important is that their intrinsic
properties may be substantially modified in a controllable way by
doping with extrinsic impurity atoms, molecules and
compounds~\cite{Duclaux}.~Recent successful experiments on
encapsulation of single atoms into single-wall CNs~\cite{Jeong}
and their intercalation into single-wall CN
bundles~\cite{Duclaux,Shimoda}, along with numerous studies of
monoatomic gas absorbtion by the CN bundles (see~\cite{Calbi} for
a review), open routes for new challenging nanophotonics
applications of atomically doped CN systems as various sources of
coherent light emitted by dopant atoms.~This, in turn, stimulates
an in-depth theoretical analysis of near-field electrodynamical
properties of atomically doped CNs.~Of primary importance towards
the nanophotonics applications are the problems of atomic
spontaneous emission and atom-nanotube van der Waals (vdW)
interactions.~The problems are not only of applied but also of
fundamental interest as they shed light on the peculiarities of
the atom-electromagnetic-field interactions in low-dimensional
dispersive and absorbing surroundings.

It has long been recognized that the spontaneous emission rate of
an excited atom is not an immutable property, but that it can be
modified by the atomic environment.~Generally called the Purcell
effect~\cite{Purcell}, the phenomenon is qualitatively explained
by the fact that the local environment modifies the strength and
distribution of the vacuum electromagnetic modes with which the
atom can interact, resulting indirectly in the alteration of
atomic spontaneous emission properties.~The Purcell effect took on
special significance recently in view of rapid progress in physics
of nanostructures. Here, the control of spontaneous emission has
been predicted to have a lot of useful applications, ranging from
the improvement of existing devices (lasers, light emitting
diodes) to such nontrivial functions as the emission of
nonclassical states of light~\cite{Weisbuch}.~In particular, the
enhancement of the spontaneous emission rate can be the first step
towards the realization of a thresholdless laser~\cite{Pelton} or
a single photon source~\cite{Vuchkovich}.~The possibility to
control atomic spontaneous emission was shown theoretically for
microcavities and microspheres~\cite{Dung,Welsch,Kimble}, optical
fibers~\cite{Klimov}, photonic crystals~\cite{John}, semiconductor
quantum dots~\cite{Sugawara}. Recent technological progress in
fabrication of low-dimensional nanostructures has enabled the
experimental investigation of spontaneous emission of single
atoms~\cite{Schniepp} and semiconductor quantum
dots~\cite{Reithmaier,Gayral,Peter} in microcavities, single
molecules in photonic crystals~\cite{Petrov}, quantum dots in
photonic crystal nanocavities~\cite{Yoshie,Finley}.

Typically, there may be two qualitatively different regimes of
interaction of an excited atomic state with a vacuum
electromagnetic field in the vicinity of the~CN. They are the weak
coupling regime and the strong coupling regime~\cite{Eberly}.~The
former is characterized by the monotonous exponential decay
dynamics of the upper atomic state with the decay rate altered
compared with the free-space value.~The latter is, in contrast,
strictly non-exponential and~is characterized by reversible Rabi
oscillations where the energy of the initially excited atom is
periodically exchanged between the atom and the~field.~A
theoretical analysis recently
done~\cite{Bondarev02,Bondarev04PLA,Bondarev04PRB} of spontaneous
decay dynamics of excited atomic states near CNs has brought out
fascinating peculiarities of the vacuum-field interactions in
atomically doped CNs.~The relative density of photonic states
(DOS) near a CN effectively increases due to the presence of
additional surface photonic states coupled with CN electronic
quasiparticle excitations~\cite{Bondarev02}. This causes an
atom-vacuum-field coupling constant, which is proportional to the
photonic DOS, to be very sensitive to the atom-CN-surface
distance. If the atom is close enough to the CN surface and the
atomic transition frequency is in the vicinity of a resonance of
the photonic DOS, the system shows strong atom-vacuum-field
coupling giving rise to rearrangement ("dressing") of atomic
levels by the vacuum-field
interaction~\cite{Bondarev04PLA,Bondarev04PRB}.~If the atom moves
away from the CN surface, the atom-field coupling strength
decreases, smoothly approaching the weak coupling regime at large
atom-surface distances since the role of surface photonic states
diminishes, causing the relative photonic DOS to decrease.~This
suggests strictly nonlinear atom-field coupling and a primary role
of the distance-dependent (local) photonic DOS in the vicinity of
the CN, so that conventional (weak-coupling) atom-field
interaction models based upon standard vacuum quantum
electrodynamics (QED) as well as those using the linear response
theory (see~\cite{Buhmann04JOB,Buhmann04pre} for a review) are, in
general, inapplicable for an atomic system in a close vicinity of
a carbon nanotube.

The aforesaid is extremely important for a proper understanding
the monoatomic gas absorbtion processes by the CNs, as they are
those arising from the atom-nanotube vdW interaction, which, from
a fundamental point of view, is the result of electrodynamic
coupling between the polarization states of the interacting
entities that originates from the vacuum fluctuations of an
electromagnetic field.~Modern {\it ab-initio} methods based on the
density functional theory (DFT) let one study the cohesive and
adsorption properties of single-wall CNs and CN bundles to an
adequate level of accuracy~\cite{Zhao2002}.~For example, recent
hydrogen~\cite{Zhang2003,Han2004} and oxygen~\cite{Dag2003}
adsorption DFT-based studies explore the CN gas-storage capacity
and the nature of hole doping processes of the semiconducting CNs
under oxygenation.~However, DFT is known to be reliable in
describing \emph{short-range} electron correlation effects whereas
the vdW energy is contributed by both short-range and long-range
interactions~\cite{Girifalco}.~The short-range contribution
consists of the repulsive part and attractive part coming,
respectively, from the overlap of core electrons on adjacent
molecules and from the decrease in electron kinetic energy due to
delocalization.~The long-range contribution is known as the London
energy~\cite{London} which originates from the (attractive)
dispersion interaction of fluctuating dipoles and dominates the
short-range contributions at distances larger than the equilibrium
one.~While adequate in describing the first two contributions, DFT
may fail in reproducing the long-range dispersion forces
correctly, especially in graphitic structures as discussed in
Ref.~\cite{Girifalco}.~An alternative (classical) approach for
computing the London dispersion energy is by summing up the
empirical $r^{-6}$ pair potentials with parameters fitted to
experimental data, or for graphitic structures often taken from
highly oriented pyrolytic
graphite~\cite{Williams2000,Ulbricht2002}.

We develop a quantum theory of near-field electrodynamical
properties of atomically doped CNs and apply it to the study of
atomic spontaneous emission and London-type dispersion
atom-nanotube interaction (simply called the vdW interaction
throughout this Chapter).~Our theory is based on a unified
macroscopic QED formalism developed for dispersing and absorbing
media in Ref.~\cite{Dung,Welsch}.~The formalism generalizes the
standard macroscopic QED normal-mode technique by representing
mode expansion in terms of a Green tensor of the (operator)
Maxwell equations in which material dispersion and absorbtion are
automatically included. In more detail, the Fourier-images of
electric and magnetic fields are considered as quantum mechanical
observables of corresponding electric and magnetic field
operators. The latter ones satisfy the Fourier-domain operator
Maxwell equations modified by the presence of a~so-called operator
noise current written in terms of a bosonic field operator and an
imaginary dielectric permittivity.~This operator is responsible
for correct commutation relations of the electric and magnetic
field operators in the presence of medium-induced absorbtion.~The
electric and magnetic field operators are then expressed in terms
of a continuum set of operator bosonic fields by means of the
convolution of the operator noise current with the electromagnetic
field Green tensor of the system.~The bosonic field operators
create and annihilate single-quantum electromagnetic medium
excitations.~They are defined by their commutation relations and
play the role of the fundamental dynamical variables in terms of
which the Hamiltonian of the composed system "electromagnetic
field + dissipative medium" is written in a standard secondly
quantized form.

The Chapter is organized as follows.~In
Section~\ref{quantization}, using the approach summarized above,
we describe the general electromagnetic field quantization scheme
and derive the total Hamiltonian for an atomic system (an atom or
a molecule) nearby a~single-wall carbon nanotube.~The nanotube is
considered as an infinitely long, infinitely thin, anisotropically
conducting cylinder.~Its surface conductivity is represented in
terms of the $\pi$-electron dispersion law obtained in the
tight-binding approximation with allowance made for the azimuthal
electron momentum quantization and the axial electron momentum
relaxation~\cite{Slepyan}.~Only the axial conductivity is taken
into account and the azimuthal one, being strongly suppressed by
transverse depolarization
fields~\cite{Benedict,Tasaki,Jorio,Li,Marinop}, is neglected.~In
Section~\ref{two-level}, based on the total Hamiltonian obtained
in Section~\ref{quantization}, we derive the simplified secondly
quantized Hamiltonian representing the "atom--nanotube" coupled
system in terms of two standard approximations.~They are the
electric dipole approximation and the atomic two-level
approximation.~This Hamiltonian generalizes that earlier applied
for the atomic spontaneous decay analysis near spherical
microcavities~\cite{Dung} and
nanotubes~\cite{Bondarev04PLA,Bondarev04PRB} in that the rotating
wave approximation is not used and the diamagnetic term of the
atom-field interaction is not neglected.~This Hamiltonian is a
starting point to derive an evolution equation for the population
probability of the upper state (Section~\ref{spontaneousdecay})
and a van der Waals energy equation of the lower (ground) state of
the two-level atomic system (Section~\ref{vdwenergy}) coupled with
the CN modified vacuum electromagnetic field.~The equations are
represented in terms of the local photonic DOS and are valid for
both strong and weak atom-field coupling. In
Section~\ref{spontaneousdecay}, we demonstrate that when the atom
is close enough to the nanotube surface and the atomic transition
frequency is in the vicinity of a resonance of the local photonic
DOS, the upper state of the coupled system decays via underdamped
Rabi oscillations which are a principal signature of strong
atom-vacuum-field coupling -- the result recently detected for
quasi-two-dimensional (2D) excitonic~\cite{Weisbuch1},
intersubband electronic~\cite{Sorba} and quasi-0D
excitonic~\cite{Reithmaier,Peter} transitions in semiconductor
quantum microcavities as well as for quasi-0D excitonic
transitions in photonic crystal nanocavities~\cite{Yoshie}.~In
Section~\ref{vdwenergy}, to take strong atom-field coupling into
account in a~correct way and thereby to advance in physical
understanding the atom-nanotube vdW interactions deeper than
simple analyzing empirical model potentials~\cite{Girifalco}, we
develop a~universal quantum mechanical approach to the
atom--nanotube vdW energy calculation.~The approach was first
suggested in Refs.~\cite{Bondarev04SSC,Bondarev05PRB}. It is based
on the perturbation theory for degenerate atomic levels (see,
e.g.,~\cite{Davydov}), thereby allowing one to account for both
strong and weak atom-vacuum-field coupling regimes in a simplest
way.~Within this approach, the ground-state atom vdW energy is
given by the integral equation which reproduces a well-known
perturbation theory result in the weak coupling regime.~By solving
it numerically, we demonstrate the inapplicability of conventional
weak-coupling-based vdW interaction models in a close vicinity of
the nanotube surface.~A summary and conclusions are given in
Section~\ref{conclusion}.

\section{Field Quantization Formalism}\label{quantization}

The quantum theory of the near-field electrodynamical properties
of carbon nanotubes involves an electromagnetic field quantization
procedure in the presence of dispersing and absorbing
bodies.~Such~a procedure faces difficulties similar to those in
quantum optics of 3D Kramers-Kronig dielectric media where the
canonical quantization scheme commonly used does not work since,
because of absorption, corresponding operator Maxwell equations
become non-Hermitian~\cite{Vogel}.~As a consequence, their
solutions cannot be expanded in power orthogonal modes and the
concept of modes itself becomes more subtle.~We, therefore, use a
unified macroscopic QED approach developed in
Refs.~\cite{Dung,Welsch}.~In this approach, the Fourier-images of
electric and magnetic fields are considered as quantum mechanical
observables of corresponding electric and magnetic field
operators.~The latter ones satisfy the Fourier-domain operator
Maxwell equations modified by the presence of a so-called operator
noise current density
$\underline{\hat{\mathbf{J}}\!}\,(\mathbf{r},\omega)$ written in
terms of a 3D vector bosonic field operator
$\hat{\mathbf{f}}(\mathbf{r},\omega)$ and a medium dielectric
tensor $\bm\epsilon(\mathbf{r},\omega)$ (supposed to be diagonal)
as\footnote{We use Gaussian units. Conversion tables to SI can be
found elsewhere (see, e.g.,~\cite{Jackson}).}
\begin{equation}
{\underline{\hat{J}\!}_{\,i}}(\mathbf{r},\omega)=\frac{\omega}{2\pi}
\sqrt{\hbar\,\mbox{Im}\epsilon_{ii}(\mathbf{r},\omega)}\,
\hat{f}_{i}(\mathbf{r},\omega),\hskip0.1cm i=1,2,3.
\label{current}
\end{equation}
This operator is responsible for correct commutation relations of
the electric and magnetic field operators in the presence of
medium-induced absorbtion.~In this formalism, the electric and
magnetic field operators are expressed in terms of a continuum set
of the 3D vector bosonic fields
$\hat{\mathbf{f}}(\mathbf{r},\omega)$ by means of the convolution
over~$\mathbf{r}$ of the current (\ref{current}) with the
classical electromagnetic field Green tensor of the system.~The
bosonic field operators
$\hat{\mathbf{f}}^{\dag}(\mathbf{r},\omega)$ and
$\hat{\mathbf{f}}(\mathbf{r},\omega)$ create and annihilate
single-quantum electromagnetic medium excitations.~They are
defined by their commutation relations
\begin{eqnarray}
\mbox{[}\,\hat{f}_{i}(\mathbf{r},\omega),
\hat{f}^{\dag}_{j}(\mathbf{r}^{\prime},\omega^{\prime})\,\mbox{]}\!\!\!&=&\!\!\!
\delta_{ij}\delta(\mathbf{r}-\mathbf{r}^{\prime})\delta(\omega-\omega^{\prime})\,,
\label{commut}\\
\mbox{[}\,\hat{f}_{i}(\mathbf{r},\omega),
\hat{f}_{j}(\mathbf{r}^{\prime},\omega^{\prime})\,\mbox{]}\!\!\!&=&\!\!\!
\mbox{[}\,\hat{f}^{\dag}_{i}(\mathbf{r},\omega),
\hat{f}^{\dag}_{j}(\mathbf{r}^{\prime},\omega^{\prime})\,\mbox{]}\,=\,0\nonumber
\end{eqnarray}
and play the role of the fundamental dynamical variables in terms
of which the Hamiltonian of the composed system "electromagnetic
field + dissipative medium" is written in a standard secondly
quantized form as
\begin{equation}
\hat{H}=\int_{0}^{\infty}\!\!\!d\omega\,\hbar\omega\!\int\!d\mathbf{r}\,
\hat{\mathbf{f}}^{\dag}(\mathbf{r},\omega)\!
\cdot\!\hat{\mathbf{f}}(\mathbf{r},\omega)\,. \label{H_field}
\end{equation}

%%%%%%%%%%%%%%%%%%%%%%%%%%%%%%%%%%%%%%%%%%%%%%%%%%%%%%%%%%%%%%%%%%%%%%%%%
\begin{figure}[t]
\begin{center}
\begin{minipage}[t]{80mm}
\epsfig{file=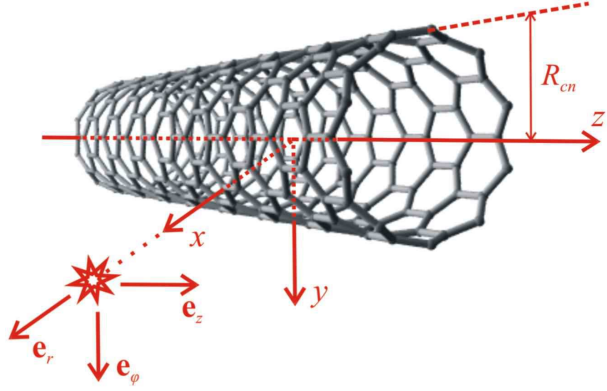, width=80mm}
\end{minipage}
\caption{The geometry of the problem.} \label{fig1}
\end{center}
\end{figure}
%%%%%%%%%%%%%%%%%%%%%%%%%%%%%%%%%%%%%%%%%%%%%%%%%%%%%%%%%%%%%%%%%%%%%%%%%%

Consider a neutral atomic system (an atom or a molecule) with its
centre-of-mass positioned at the point $\mathbf{r}_{A}$ near an
infinitely long single-wall CN. Since the problem has a cylindric
symmetry, assign the orthonormal cylindric basis
$\{\mathbf{e}_{r},\mathbf{e}_{\varphi},\mathbf{e}_{z}\}$ with
$\mathbf{e}_{z}$ directed along the nanotube axis and
$\mathbf{r}_{A}=r_{A}\mathbf{e}_{r}=\{r_{A},0,0\}$~(see
Fig.~\ref{fig1}). For carbon nanotubes, their strong transverse
depolarization along with transverse electron momentum
quantization allow one to neglect the azimuthal current and radial
polarizability~\cite{Slepyan,Benedict,Tasaki,Jorio,Li,Marinop}, in
which case the dielectric tensor components $\epsilon_{rr}$ and
$\epsilon_{\varphi\varphi}$ are identically equal to unit.~The
component $\epsilon_{zz}$ is caused by the CN longitudinal
polarizability and is responsible for the axial surface current
parallel to the $\textbf{e}_{z}$ vector.~This current may be
represented in terms of the 1D bosonic field operators by analogy
with Eq.~(\ref{current}).~Indeed, taking into account the
dimensionality conservation in passing from bulk to a~monolayer in
Eq.~(\ref{H_field}) and using a simple Drude
relation~\cite{Tasaki}
\begin{equation}
\sigma_{zz}(\mathbf{R},\omega)=
-i\omega\frac{\epsilon_{zz}(\mathbf{R},\omega)-1}{4\pi
S\rho_{T}}\,, \label{sigmaCN}
\end{equation}
where $\mathbf{R}=\!\{R_{cn},\phi,Z\}$~is the radius-vector of an
arbitrary point of the CN surface, $R_{cn}$ is the radius of the
CN, $\sigma_{zz}(\mathbf{R},\omega)$ is the CN surface axial
conductivity per unit length, $S$ is the area of a single
nanotube, $\rho_{T}$ is the cubic density of the tubule, one
immediately has from Eq.~(\ref{current})
\begin{equation}
\underline{\hat{\mathbf{J}}\!}\,(\mathbf{R},\omega)\!=
\!\sqrt{\frac{\hbar\omega\mbox{Re}\sigma_{zz}(\mathbf{R},\omega)}{\pi}}\,
\hat{f}(\mathbf{R},\omega)\textbf{e}_{z} \label{currentCN}
\end{equation}
with $\hat{f}(\mathbf{R},\omega)$ being the 1D bosonic field
operator defined on the CN surface.~The operators
$\hat{f}^{\dag}(\mathbf{R},\omega)$ and
$\hat{f}(\mathbf{R},\omega)$ satisfy the commutation relations
(\ref{commut}) on the surface of the CN. They create and
annihilate the single-quantum bosonic-type electromagnetic medium
excitation of the frequency $\omega$ at the point $\mathbf{R}$ of
the CN surface.

We are now in a position to define the Hamiltonian of the
system.~Following the general macroscopic QED approach in the
presence of absorbing and dispersive media~\cite{Welsch}, we
introduce the total nonrelativistic Hamiltonian of an atomic
system interacting with a CN modified quantized vacuum
electromagnetic field in the form
\begin{eqnarray}
\hat{H}\!\!\!&=&\!\!\!\int_{0}^{\infty}\!\!\!\!\!d\omega\,\hbar\omega\!\int\!d\mathbf{R}
\,\hat{f}^{\dag}(\mathbf{R},\omega)\hat{f}(\mathbf{R},\omega)\label{Htotini}\\
&+&\!\!\!\sum_{i}\frac{1}{2m_{i}}\left[\hat{\mathbf{p}}_{i}-
\frac{q_{i}}{c}\hat{\mathbf{A}}(\mathbf{r}_{A}+\hat{\mathbf{r}}_{i})\right]^{2}
+\frac{1}{2}\int\!d\mathbf{r}\,\hat{\rho}_{A}(\mathbf{r})\hat{\varphi}_{A}(\mathbf{r})
+\int\!d\mathbf{r}\,\hat{\rho}_{A}(\mathbf{r})\hat{\varphi}(\mathbf{r})\,,\nonumber
\end{eqnarray}
where $m_{i}$, $q_{i}$, $\hat{\mathbf{r}}_{i}$ and
$\hat{\mathbf{p}}_{i}$ are, respectively, the masses, charges,
coordinates (relative to $\mathbf{r}_{A}$) and momenta of the
particles constituting the atomic subsystem.~The first term of the
Hamiltonian describes the quantized \emph{medium-assisted} vacuum
electromagnetic field in the presence of the CN.~The second and
the third terms represent the kinetic energy of the charged
particles and their mutual Coulomb interaction, respectively, with
\begin{equation}
\hat{\varphi}_{A}(\mathbf{r})=\int\!d\mathbf{r}^{\prime}
\frac{\hat{\rho}_{A}(\mathbf{r}^{\prime})}{|\mathbf{r}-\mathbf{r}^{\prime}|}
\label{phia}
\end{equation}
being the scalar potential of the charged particles distributed
with the density
$\hat{\rho}_{A}(\mathbf{r})=\sum_{i}q_{i}\delta(\mathbf{r}-\mathbf{r}_{A}-\hat{\mathbf{r}}_{i})$
in the atomic subsystem.~The last term of the Hamiltonian accounts
for the Coulomb interaction of the particles with the CN. The
vector potential $\hat{\mathbf{A}}$ and the scalar potential
$\hat{\varphi}$ of the CN modified electromagnetic field are given
for an arbitrary $\mathbf{r}=\{r,\varphi,z\}$ in the
Schr\"{o}dinger picture by
\begin{equation}
\hat{\mathbf{A}}(\mathbf{r})=\int_{0}^{\infty}\!\!\!\!\!d\omega\frac{c}{i\omega}\,
\underline{\hat{\mathbf{E}}}^{\perp}(\mathbf{r},\omega)+\mbox{h.c.},
\label{vecpot}
\end{equation}
\begin{equation}
-\bm{\nabla}\hat{\varphi}(\mathbf{r})=\int_{0}^{\infty}\!\!\!\!\!d\omega\,
\underline{\hat{\mathbf{E}}}^{\parallel}(\mathbf{r},\omega)+\mbox{h.c.},
\label{scalpot}
\end{equation}
where
\begin{equation}
\underline{\hat{\mathbf{E}}}^{\perp(\parallel)}(\mathbf{r},\omega)=
\int\!d\mathbf{r}^{\prime}\,\bm{\delta}^{\perp(\parallel)}
(\mathbf{r}-\mathbf{r}^{\prime})\cdot\underline{\hat{\mathbf{E}}}
(\mathbf{r}^{\prime},\omega) \label{elperpar}
\end{equation}
is the transverse (longitudinal) electric field operator with
\begin{equation}
\delta^{\parallel}_{\alpha\beta}(\mathbf{r})=
-\nabla_{\alpha}\nabla_{\beta}\frac{1}{4\pi r}
\label{deltapar}
\end{equation}
and
\begin{equation}
\delta^{\perp}_{\alpha\beta}(\mathbf{r})=\delta_{\alpha\beta}\,
\delta(\mathbf{r})-\delta^{\parallel}_{\alpha\beta}(\mathbf{r})
\label{deltaper}
\end{equation}
being the longitudinal and transverse dyadic $\delta$-functions,
respectively, and $\underline{\hat{\mathbf{E}}}$ representing the
total electric field operator which satisfies the following set of
Fourier-domain Maxwell equations
\begin{equation}
\bm{\nabla}\times\underline{\hat{\mathbf{E}}}(\mathbf{r},\omega)=
ik\,\underline{\hat{\mathbf{H}}}(\mathbf{r},\omega),
\label{MaxwelE}
\end{equation}
\begin{equation}
\bm{\nabla}\times\underline{\hat{\mathbf{H}}}(\mathbf{r},\omega)=
-ik\,\underline{\hat{\mathbf{E}}}(\mathbf{r},\omega)+
\frac{4\pi}{c}\,\underline{\hat{\mathbf{I}}}(\mathbf{r},\omega).
\label{MaxwelH}
\end{equation}
Here, $\underline{\hat{\mathbf{H}}}$ stands for the magnetic field
operator, $k=\omega/c$, and
\begin{equation}
\underline{\hat{\mathbf{I}}}(\mathbf{r},\omega)=\int\!\!d\mathbf{R}\,
\delta(\mathbf{r}-\mathbf{R})\,\underline{\hat{\mathbf{J}}\!}\,(\mathbf{R},\omega)
=2\underline{\hat{\mathbf{J}}\!}\,(R_{cn},\varphi,z,\omega)\,\delta(r-R_{cn})\,,
\label{Irw}
\end{equation}
is the exterior operator current density [with
$\underline{\hat{\mathbf{J}}}(\mathbf{R},\omega)$ defined by
Eq.~(\ref{currentCN})] associated with the presence of the CN.

From Eqs.~(\ref{MaxwelE})-(\ref{Irw}) in view of
Eq.~(\ref{currentCN}), it follows that
\begin{equation}
\underline{\hat{\mathbf{E}}}(\mathbf{r},\omega)=
i\frac{4\pi}{c}\,k\!\int\!\!d\mathbf{R}\,\mathbf{G}(\mathbf{r},\mathbf{R},\omega)
\!\cdot\!\underline{\hat{\mathbf{J}}\!}\,(\mathbf{R},\omega)
\label{Erw}
\end{equation}
[and $\underline{\hat{\mathbf{H}}}=(ik)^{-1}
\bm{\nabla}\times\underline{\hat{\mathbf{E}}}$ accordingly], where
$\mathbf{G}$ is the Green tensor of the classical electromagnetic
field in the vicinity of the CN.~The Green tensor components
satisfy the equation
\begin{equation}
\sum_{\alpha=r,\varphi,z}\!\!\!
\left(\bm{\nabla}\!\times\bm{\nabla}\!\times-\,k^{2}\right)_{\!z\alpha}
G_{\alpha z}(\mathbf{r},\mathbf{R},\omega)=
\delta(\mathbf{r}-\mathbf{R}), \label{GreenequCN}
\end{equation}
together with the radiation conditions at infinity and the
boundary conditions on the CN surface.~The set of
Eqs.~(\ref{Htotini})-(\ref{GreenequCN}) forms a closed
electromagnetic field quantization formalism in the presence of
dispersing and absorbing media which meets all the basic
requirements of the~standard quantum
electrodynamics~\cite{Welsch}.~All the information about medium,
which is a carbon nanotube in our case, is contained in the Green
tensor $\mathbf{G}$ via the CN surface axial conductivity
$\sigma_{zz}$ in Eq.~(\ref{currentCN}) which comes into play when
$\mathbf{G}$ is imposed the boundary conditions on the CN surface.
The classical electromagnetic field Green tensor of the
"atom--nanotube" system is derived in Appendix~B.

Assuming further that the atomic subsystem is sufficiently
localized in space, so that the long-wavelength approximation
applies, one can expand the field operators
$\hat{\mathbf{A}}(\mathbf{r})$ and $\hat{\varphi}(\mathbf{r})$ in
the Hamiltonian (\ref{Htotini}) around the atomic center of mass
position $\mathbf{r}_{A}$ and only keep the leading non-vanishing
terms of the expansions.~Then, under the condition of the Coulomb
gauge $[\hat{\mathbf{p}}_{i},\hat{\mathbf{A}}]=0$, one arrives at
the approximate total Hamiltonian of the form
\begin{equation}
\hat{H}=\hat{H}_{F}+\hat{H}_{A}+\hat{H}^{(1)}_{AF}+\hat{H}^{(2)}_{AF}\,,
\label{Htot}
\end{equation}
where
\begin{equation}
\hat{H}_{F}=\int_{0}^{\infty}\!\!\!\!\!d\omega\,\hbar\omega\!\int\!d\mathbf{R}
\,\hat{f}^{\dag}(\mathbf{R},\omega)\hat{f}(\mathbf{R},\omega)\,,
\label{Hf}
\end{equation}
\begin{equation}
\hat{H}_{A}=\sum_{i}\frac{\hat{\mathbf{p}}^{2}}{2m_{i}}+
\sum_{i<j}\frac{q_{i}q_{j}}{|\mathbf{r}_{i}-\mathbf{r}_{j}|}\,,
\label{Ha}
\end{equation}
\begin{equation}
\hat{H}^{(1)}_{AF}=-\sum_{i}\frac{q_{i}}{m_{i}c}\,\hat{\mathbf{p}}_{i}
\!\cdot\!\hat{\mathbf{A}}(\mathbf{r}_{A})+\hat{\mathbf{d}}\!\cdot\!
\bm{\nabla}\hat{\varphi}(\mathbf{r}_{A})\,, \label{Haf1}
\end{equation}
\begin{equation}
\hat{H}^{(2)}_{AF}=\sum_{i}\frac{q^{2}_{i}}{2m_{i}c^{2}}\,
\hat{\mathbf{A}}^{\!2}(\mathbf{r}_{A}) \label{Haf2}
\end{equation}
are, respectively, the Hamiltonian of the vacuum electromagnetic
field modified by the presence of the CN, the Hamiltonian of the
atomic subsystem, and the electric dipole approximation for the
Hamiltonian of the atom-field interaction (separated into two
contributions according to their role in the vdW energy -- see
below) with $\hat{\mathbf{d}}=\sum_{i}q_{i}\hat{\mathbf{r}}_{i}$
being the electric dipole moment operator of the atomic subsystem.

\section{Total Hamiltonian in terms of the local photonic DOS}\label{two-level}

Starting from Eqs.~(\ref{Htot})-(\ref{Haf2}) and using
Eqs.~(\ref{vecpot})-(\ref{deltaper}), (\ref{Erw}) and
(\ref{currentCN}), one obtains the following two-level
approximation for the total Hamiltonian of the system (see
Appendix~A)
\begin{equation}
\hat{\cal{H}}=\hat{\cal{H}}_{0}+\hat{\cal{H}}_{int}\,,
\label{Htwolev}
\end{equation}
where the first term stands for the unperturbed Hamiltonian given
by
\begin{equation}
\hat{\cal{H}}_{0}=\hat{H}_{F}+\hat{\cal{H}}_{A} \label{H0}
\end{equation}
with $\hat{H}_{F}$ being the Hamiltonian (\ref{Hf}) of the CN
modified field and
\begin{equation}
\hat{\cal{H}}_{A}=\frac{\hbar\tilde{\omega}_{A}}{2}\,\hat{\sigma}_{z}
\label{Harenorm}
\end{equation}
representing the 'effective' unperturbed two-level atomic
subsystem, and the second term stands for their interaction given
by
\begin{equation}
\hat{\cal{H}}_{int}=\int_{0}^{\infty}\!\!\!\!\!d\omega\!\int\!d\mathbf{R}\,[\,
\mbox{g}^{(+)}(\mathbf{r}_{A},\mathbf{R},\omega)\,\hat{\sigma}^{\dag}
-\mbox{g}^{(-)}(\mathbf{r}_{A},\mathbf{R},\omega)\,\hat{\sigma}\,]\,
\hat{f}(\mathbf{R},\omega)+\mbox{h.c.}. \label{Hint}
\end{equation}
Here, the Pauli operators
\begin{eqnarray}
\hat{\sigma}=|l\rangle\langle u|\,,\hskip0.2cm
\hat{\sigma}^{\dag}=|u\rangle\langle l|\,,\hskip0.2cm
\hat{\sigma}_{z}&\!=\!&|u\rangle\langle u|-|l\rangle\langle l|\,,\label{Pauli}\\
|u\rangle\langle u|+|l\rangle\langle l|&\!=\!&\hat{I}\,,\nonumber
\end{eqnarray}
describe electric dipole transitions between the two ato\-mic
states, upper $|u\rangle$ and lower $|l\rangle$, separated by the
transition frequency $\omega_{A}$.~This 'bare' transition
frequency is modified by the interaction (\ref{Haf2}) which, being
independent of the atomic dipole moment, does not contribute to
mixing the $|u\rangle$ and $|l\rangle$ states, giving rise,
however, to the new \emph{renormalized} transition frequency
\begin{equation}
\tilde{\omega}_{A}=\omega_{A}\!\left[1-\frac{2}{(\hbar\omega_{A})^{2}}
\int_{0}^{\infty}\!\!\!\!\!d\omega\!\int\!d\mathbf{R}\,
|\mbox{g}^{\perp}(\mathbf{r}_{A},\mathbf{R},\omega)|^{2}\right]
\label{omegarenorm}
\end{equation}
in the 'effective' atomic Hamiltonian~(\ref{Harenorm}).~On the
contrary, the interaction (\ref{Haf1}), being dipole moment
dependent, mixes the $|u\rangle$ and $|l\rangle$ states, yielding
the perturbation Hamiltonian~(\ref{Hint}) with the (dipole)
interaction matrix elements of the form
\begin{equation}
\mbox{g}^{(\pm)}(\mathbf{r}_{A},\mathbf{R},\omega)=
\mbox{g}^{\perp}(\mathbf{r}_{A},\mathbf{R},\omega)\pm
\frac{\omega}{\omega_{A}}\,\mbox{g}^{\parallel}(\mathbf{r}_{A},\mathbf{R},\omega)\,,
\label{gpm}
\end{equation}
where
\begin{equation}
\mbox{g}^{\perp(\parallel)}(\mathbf{r}_{A},\mathbf{R},\omega)=
-i\frac{4\omega_{A}}{c^{2}}\,d_{z}\sqrt{\pi\hbar\omega\,\mbox{Re}\,
\sigma_{zz}(\mathbf{R},\omega)}\;^{\perp(\parallel)}
G_{zz}(\mathbf{r}_{A},\mathbf{R},\omega) \label{gperppar}
\end{equation}
with $d_{z}\!=\!\langle l|\hat{d}_{z}|u\rangle\!=\!\langle
u|\hat{d}_{z}|l\rangle$ being the matrix element of the atomic
dipole moment $z$-component and
\begin{equation}
^{\perp(\parallel)}G_{zz}(\mathbf{r}_{A},\mathbf{R},\omega)=\int\!d\mathbf{r}\,
\delta^{\perp(\parallel)}_{zz}(\mathbf{r}_{A}-\mathbf{r})\,
G_{zz}(\mathbf{r},\mathbf{R},\omega) \label{Gzzperpar}
\end{equation}
representing the field Green tensor $zz$-component transverse
(longitudinal) with respect to the first variable.~This is the
only Green tensor component we have to take account of.~All the
other components can be safely neglected, because our model
neglects the azimuthal current and radial polarizability of the CN
from the very outset [see the discussion above
Eq.~(\ref{sigmaCN})].

Matrix elements (\ref{gpm})~and~(\ref{gperppar}) have the
following properties (Appendix~B)
\begin{equation}
\int\!d\mathbf{R}\,|\mbox{g}^{\perp(\parallel)}(\mathbf{r}_{A},\mathbf{R},\omega)|^{2}\!=
\frac{(\hbar\omega_{A})^{2}}{2\pi\omega^{2}}\,\Gamma_{0}(\omega)\,
\xi^{\perp(\parallel)}(\mathbf{r}_{A},\omega) \label{DOS}
\end{equation}
and
\begin{equation}
\int\!d\mathbf{R}\,|\mbox{g}^{(\pm)}(\mathbf{r}_{A},\mathbf{R},\omega)|^{2}=
\frac{(\hbar\omega_{A})^{2}}{2\pi\omega^{2}}\,\Gamma_{0}(\omega)\!\left[
\xi^{\perp}(\mathbf{r}_{A},\omega)+\!\left(\frac{\omega}{\omega_{A}}\right)^{\!2}
\xi^{\parallel}(\mathbf{r}_{A},\omega)\right]. \label{DOSparper}
\end{equation}
Here, $\xi^{\perp}(\mathbf{r}_{A},\omega)$ and
$\xi^{\parallel}(\mathbf{r}_{A},\omega)$ are the transverse and
longitudinal \emph{local} (position-dependent) photonic DOS
functions defined by
\begin{equation}
\xi^{\perp(\parallel)}(\mathbf{r}_{A},\omega)=
\frac{\mbox{Im}^{\perp(\parallel)}G_{zz}^{\perp(\parallel)}(\mathbf{r}_{A},\mathbf{r}_{A},\omega)}
{\mbox{Im}\,G_{zz}^{0}(\mathbf{r}_{A},\mathbf{r}_{A},\omega)}\,,
\label{DOSdef}
\end{equation}
where
\begin{equation}
^{\perp(\parallel)}G_{zz}^{\perp(\parallel)}(\mathbf{r}_{A},\mathbf{r}_{A},\omega)=
\!\int\!d\mathbf{r}\,d\mathbf{r}^{\prime}\,\delta^{\perp(\parallel)}_{zz}
(\mathbf{r}_{A}-\mathbf{r})\;G_{zz}(\mathbf{r},\mathbf{r}^{\prime},\omega)\,
\delta^{\perp(\parallel)}_{zz}(\mathbf{r}^{\prime}-\mathbf{r}_{A})
\label{Gzzperparperpar}
\end{equation}
is the Green tensor $zz$-component transverse (longitudinal) with
respect to both coordinate variables,
\begin{equation}
\mbox{Im}\,G^{0}_{zz}(\mathbf{r}_{A},\mathbf{r}_{A},\omega)=
\frac{\omega}{6\pi c} \label{G0}
\end{equation}
is the vacuum imaginary Green tensor
$zz$-component~\cite{Abrikosov}, and
\begin{equation}
\Gamma_{0}(\omega)=\frac{8\pi d_{z}^{2}\omega^{2}}{3\hbar c^{2}}\,
\mbox{Im}\,G^{0}_{zz}(\mathbf{r}_{A},\mathbf{r}_{A},\omega)=
\frac{4d_{z}^{2}\omega^{3}}{3\hbar c^{3}} \label{Gamma0}
\end{equation}
is the atomic spontaneous decay rate for the $z$-oriented atomic
dipole in vacuum, taken at an arbitrary frequency
$\omega$~\cite{Barnett,Agarwal75}.

Note that the transverse and longitudinal imaginary Green tensor
components in Eq.~(\ref{DOSdef}) can be written as
\begin{eqnarray}
\mbox{Im}^{\perp}G_{zz}^{\perp}\!\!\!&=&\!\!\!\mbox{Im}\,G^{0}_{zz}+\mbox{Im}^{\perp}
\overline{G}_{zz}^{\,\perp}\,,\label{ImGzzper}\\
\mbox{Im}^{\,\parallel}G_{zz}^{\parallel}\!\!\!&=&\!\!\!\mbox{Im}^{\,\parallel}
\overline{G}_{zz}^{\,\parallel} \label{ImGzzppar}
\end{eqnarray}
with $^{\perp(\parallel)}\overline{G}_{zz}^{\perp(\parallel)}$
representing the "pure" CN contribution to the total imaginary
Green tensor.~The longitudinal imaginary Green tensor in
Eq.~(\ref{ImGzzppar}) is totally contributed by a (longitudinal)
CN static polarization field and, therefore, does not contain the
vacuum term which is the transverse one by its definition as there
are no polarization Coulomb sources in
vacuum~\cite{Jackson,Abrikosov}.~Thus, Eq.~(\ref{DOSdef}) may, in
view of Eqs.~(\ref{ImGzzper}) and (\ref{ImGzzppar}), be rewritten
in the form
\begin{eqnarray}
\xi^{\perp}(\mathbf{r}_{A},\omega)\!\!\!&=&\!\!\!
1+\overline{\xi}^{\perp}(\mathbf{r}_{A},\omega)\,,\label{xiper}\\
\xi^{\parallel}(\mathbf{r}_{A},\omega)\!\!\!&=&\!\!\!
\overline{\xi}^{\,\parallel}(\mathbf{r}_{A},\omega)\,,\label{xipar}
\end{eqnarray}
where the $\mathbf{r}_{A}$-dependent terms come from the presence
of the CN. For $r_{A}\!>\!R_{cn}$, they are explicitly given by
(Appendix~B)
\begin{eqnarray}
\overline{\xi}^{\,\perp}(\mathbf{r}_{A},x)=\frac{3}{\pi}\,\mbox{Im}\!\!\!
\sum_{p=-\infty}^{\infty}\!\int_{C}\!\frac{dy\,s(R_{cn},x)v(y)^{4}
I_{p}^{2}[v(y)u(R_{cn})x]K_{p}^{2}[v(y)u(r_{A})x]}
{1+s(R_{cn},x)v(y)^{2}I_{p}[v(y)u(R_{cn})x]K_{p}[v(y)u(R_{cn})x]}\,,\;\;\;
\label{ksiper}\\
\overline{\xi}^{\,\parallel}(\mathbf{r}_{A},x)=\frac{3}{\pi}\,\mbox{Im}\!\!\!
\sum_{p=-\infty}^{\infty}\!\int_{C}\!\frac{dy\,s(R_{cn},x)y^{2}v(y)^{2}
I_{p}^{2}[v(y)u(R_{cn})x]K_{p}^{2}[v(y)u(r_{A})x]}
{1+s(R_{cn},x)v(y)^{2}I_{p}[v(y)u(R_{cn})x]K_{p}[v(y)u(R_{cn})x]}\,,\;\;\;\;
\label{ksipar}
\end{eqnarray}
where
\begin{equation}
x=\frac{\hbar}{2\gamma_{0}}\,\omega
\label{omegadimless}
\end{equation}
is the dimensionless frequency with $\gamma_{0}=2.7$~eV being the
carbon nearest neighbor hopping integral~\cite{Wallace} appearing
in the CN surface axial conductivity~in~Eq.~(\ref{currentCN}),
$I_{p}$ and $K_{p}$ are the modified cylindric Bessel functions,
$v(y)=\sqrt{y^{2}-1}\,$, $u(r)=2\gamma_{0}r/\hbar c$, and
$s(R_{cn},x)=2i\alpha u(R_{cn})x\overline{\sigma}_{zz}(R_{cn},x)$
with $\overline{\sigma}_{zz}=2\pi\hbar\sigma_{zz}/e^{2}$ being the
dimensionless CN surface conductivity per unit length and
$\alpha=e^{2}/\hbar c=1/137$ representing the fine-structure
constant.~The integration contour $C$ runs along the real axis of
the complex plane and envelopes the branch points $y=\pm 1$ of the
function $v(y)$ in the integrands from below and from above,
respectively.~For $r_{A}\!<\!R_{cn}$, Eqs.~(\ref{ksiper}) and
(\ref{ksipar}) are modified by a simple replacement
$r_{A}\!\leftrightarrow\!R_{cn}$ in the Bessel function arguments
in the numerators of the integrands.~Note the divergence of the
local photonic DOS functions
$\overline{\xi}^{\,\perp(\parallel)}(\mathbf{r}_{A},x)$ at
$r_{A}=R_{cn}\,$, i.~e. when the atom is located right on the CN
surface.~Mathematically, this comes from the logarithmic
divergence of the Bessel function $K_{0}$ at small arguments (see,
e.g.,~\cite{Abramovitz}), yielding the divergent product
$I_{0}K_{0}$ and, as a consequence, the divergent integrands in
Eqs.~(\ref{ksiper}) and (\ref{ksipar}), if and only if the
arguments of the functions $I_{0}$ and $K_{0}$ in the numerator of
the integrand vanish simultaneously.~Physically, the point is that
the~CN dielectric tensor longitudinal component $\epsilon_{zz}$
[which, according to Eqs.~(\ref{sigmaCN}),~(\ref{currentCN}) and
(\ref{Erw}) is responsible for the surface axial conductivity
$\sigma_{zz}$ in Eqs.~(\ref{ksiper}) and (\ref{ksipar})] is
obtained as a result of a standard procedure of \emph{physical}
averaging a~local electromagnetic field over the two spatial
directions in the graphene plane~\cite{Lambin}.~Such averaging
does not assume extrinsic atoms on the graphene surface. To take
them into consideration the averaging procedure must be modified.
Thus, the applicability domain of our model is restricted by the
condition
\begin{equation}
\mid\!r_{A}\!-R_{cn}\!\mid\,>a\,, \label{condition}
\end{equation}
where $a=1.42$~\AA~ is the graphene interatomic
distance~\cite{Wallace}.

Eqs.~(\ref{Htwolev})--(\ref{gperppar}) represent the total
secondly quantized Hamiltonian of an atomic system interacting
with the electromagnetic field in the vicinity of the carbon
nanotube.~In deriving this Hamiltonian, there were only two
standard approximations done.~They are the electric dipole
approximation and the two-level approximation. The rotating wave
approximation commonly used was not applied, and the (diamagnetic)
$\hat{\mathbf{A}}^{\!2}$-term of the atom-field interaction
[Eq.~(\ref{Haf2})] was not neglected.~The latter two
approximations are justified in analyzing an atomic state
evolution due to \emph{real} dipole transitions to the nearest
states (shown schematically in Fig.~\ref{fig2})~\cite{Eberly},
described for two-level systems by the $\mbox{g}^{(+)}$-terms of
the Hamiltonian~(\ref{Hint}).~They are, however, inappropriate in
calculating the energy, where the initial energy level is being
shifted due to the perturbation caused by \emph{virtual} dipole
transitions via intermediate atomic states (see Fig.~\ref{fig6} in
Section~\ref{vdwenergy}), so that the system always remains in the
initial state and the terms neglected contribute to the level
shift to the same order of magnitude.~For two-level systems, this
is described by $\mbox{g}^{(-)}$-terms of the
Hamiltonian~(\ref{Hint}) (usually neglected within the rotating
wave approximation) together with the atomic transition
frequency~(\ref{omegarenorm}) renormalized due to the (usually
discarded) $\hat{\mathbf{A}}^{\!2}$-term of the atom-field
interaction.

\section{Spontaneous Decay Dynamics of an Excited Atomic State near a~Carbon Nanotube}\label{spontaneousdecay}

\begin{figure}[t]
\begin{center}
\begin{minipage}[h]{100mm}
\epsfig{file=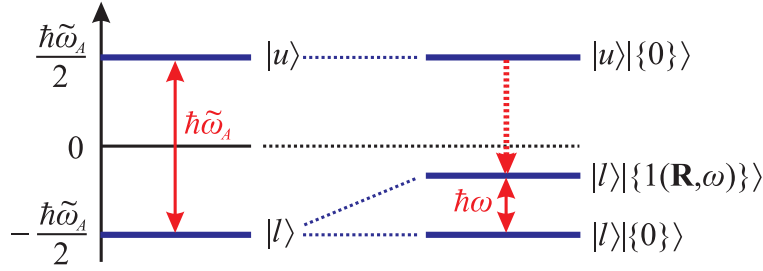, width=100mm}
\end{minipage}
\caption{Schematic of the energy levels of the "atom--nanotube"
coupled system in the problem of the spontaneous decay. On the
left are the unperturbed atomic levels given by the
Hamiltonian~(\ref{Harenorm}), on the right are the levels of the
coupled system. The relevant real transition of the system is
shown by the vertical dashed arrow. See Eqs.~(\ref{tildexA}) and
(\ref{omegadimless}) for $\tilde{\omega}_{A}$.} \label{fig2}
\end{center}
\end{figure}

When the atom is initially in the upper state and the field
subsystem is in vacuum (see Fig.~\ref{fig2}), the time-dependent
wave function of the whole system can be written~as
\begin{eqnarray}
|\psi(t)\rangle=C_{u}(t)\,e^{-i(\tilde{\omega}_{A}/2)t}|u\rangle|\{0\}\rangle
\hskip1.5cm\nonumber\\[0.3cm]+\int\!d\mathbf{R}\!\int_{0}^{\infty}\!\!\!d\omega\,
C_{l}(\mathbf{R},\omega,t)\,e^{-i(\omega-\tilde{\omega}_{A}/2)t}
|l\rangle|\{1(\mathbf{R},\omega)\}\rangle, \label{wfuncspont}
\end{eqnarray}
where $|\{0\}\rangle$ is the vacuum state of the field subsystem,
$|\{1(\mathbf{R},\omega)\}\rangle$ is its excited state where the
field is in a single-quantum Fock state, $C_{u}$ and $C_{l}$ are
the population probability amplitudes of the upper state and lower
state of the \emph{whole} system, respectively.~The function
(\ref{wfuncspont}) accounts for the lower state degeneracy with
respect to $\mathbf{R}$ as well as its possible degeneracy with
respect to $\omega$ when $\omega\sim\tilde{\omega}_{A}$ as is seen
in Fig.~\ref{fig2}.

%%%%%%%%%%%%%%%%%%%%%%%%%%%%%%%%%%%%%%%%%%%%%%%%%%%%%%%%%%%%%%%%%%%%%%%%%
%%%%%%%%%%%%%%%%%%%%%%%%%%%%%%%%%%%%%%%%%%%%%%%%%%%%%%%%%%%%%%%%%%%%%%%%%%

The time-dependent Schr\"{o}dinger equation with the
Hamiltonian~(\ref{Htwolev})-(\ref{Hint}) and the wave
function~(\ref{wfuncspont}) yields the following evolution law for
the population probability amplitude $C_{u}$ of the upper state
\begin{eqnarray}
C_{u}(t)=1+\int_{0}^{t}\!K(t-t^{\prime})\,C_{u}(t^{\prime})\,dt^{\prime},
\label{Volterra}\hskip2.5cm\\
K(t-t^{\prime})=\frac{1}{\hbar^{2}}\int_{0}^{\infty}\!\!\!d\omega\,
\frac{e^{-i(\omega-\tilde{\omega}_{A})(t-t^{\prime})}-1}
{\displaystyle i\,(\omega-\tilde{\omega}_{A})}\!
\int\!d\mathbf{R}\,|\mbox{g}^{(+)}(\mathbf{r}_{A},\mathbf{R},\omega)|^{2}.
\label{kernel}
\end{eqnarray}
This, upon recalling
Eqs.~(\ref{DOSparper}),~(\ref{xiper}),~(\ref{xipar}) and
(\ref{omegadimless}) and introducing the dimensionless time
\begin{equation}
\tau=\frac{2\gamma_{0}}{\hbar}\,t, \label{timedimless}
\end{equation}
reduces in dimensionless variables to
\begin{eqnarray}
C_{u}(\tau)=1+\int_{0}^{\tau}\!K(\tau-\tau^{\prime})\,
C_{u}(\tau^{\prime})\,d\tau^{\prime},\label{Volterradimless}\hskip2cm\\
K(\tau-\tau^{\prime})=\frac{x_{A}^{2}}{2\pi}\!\int_{0}^{\infty}\!\!\!dx\,
\frac{\tilde{\Gamma}_{0}(x)}{x^{2}}\left[
1+\overline{\xi}^{\perp}(\mathbf{r}_{A},x)+\left(\!\frac{x}{x_{A}}\!\right)^{\!2}\!
\overline{\xi}^{\,\parallel}(\mathbf{r}_{A},x)\right]\nonumber\\
\times\frac{e^{-i(x-\tilde{x}_{A})(\tau-\tau^{\prime})}-1}{i\,(x-\tilde{x}_{A})}\hskip5.4cm
\label{kerneldimless}
\end{eqnarray}
with the $\mathbf{r}_{A}$-dependent transverse and longitudinal
local photonic DOS functions
$\overline{\xi}^{\perp}(\mathbf{r}_{A},x)$ and
$\overline{\xi}^{\,\parallel}(\mathbf{r}_{A},x)$ given explicitly
by Eqs.~(\ref{ksiper}) and (\ref{ksipar}), respectively,
\begin{equation}
\tilde{\Gamma}_{0}(x)=\frac{\hbar}{2\gamma_{0}}\,\Gamma_{0}(x)=
\frac{4d_{z}^{2}}{3\hbar c^{3}}
\left(\frac{2\gamma_{0}}{\hbar}\right)^{\!2}x^{3} \label{G0x}
\end{equation}
being the dimensionless vacuum spontaneous decay rate
(\ref{Gamma0}), and
\begin{equation}
\tilde{x}_{A}=x_{A}\left[1-\frac{1}{\pi}\int_{0}^{\infty}\!\!\!dx\,
\frac{\tilde{\Gamma}_{0}(x)}{x^{2}}\,\overline{\xi}^{\perp}(\mathbf{r}_{A},x)\right]
\label{tildexA}
\end{equation}
representing the dimensionless renormalized transition frequency
(\ref{omegarenorm}) where a~divergent contribution to the vacuum
Lamb shift has been omitted since the true (renormalized) vacuum
Lamb shift may be considered as included in the atomic transition
frequency~$x_{A}$~\cite{Dung,Welsch}.

Eq.~(\ref{Volterradimless}) is a well-known Volterra integral
equation of the second kind, which in our case describes
spontaneous decay dynamics of an excited two-level atomic system
interacting with the vacuum electromagnetic field modified by the
presence of a CN. All the CN parameters that are relevant for the
spontaneous decay are contained in the local photonic DOS
functions in the kernel~(\ref{kerneldimless}). The latter ones
are, in view of
Eqs.~(\ref{DOSdef}),~(\ref{xiper})~and~(\ref{xipar}), determined
by the imaginary classical Green tensor of the CN modified
electromagnetic field.

\subsection{Qualitative Analysis}\label{spontqualitatively}

Let us first analyze qualitatively the time dynamics of the upper
state population probability $C_{u}(\tau)$ in terms of two
different approximations admitting analytical solutions of the
evolution problem~(\ref{Volterradimless}),~(\ref{kerneldimless}).
They are the Markovian approximation and the single-resonance
approximation of the local photonic DOS in the vicinity of the CN.

\subsubsection{Markovian Approximation}

Suppose the atom is faraway from the CN. Obviously, the local
photonic DOS functions
$\overline{\xi}^{\,\perp}(\mathbf{r}_{A},x)$ and
$\overline{\xi}^{\,\parallel}(\mathbf{r}_{A},x)$ are then small in
their values and the atom-field coupling matrix elements
(\ref{DOS}) and (\ref{DOSparper}) are mainly contributed by the
free space term which comes from the unity in $\xi^{\perp}$ [see
Eqs.~(\ref{DOSdef}),~(\ref{ImGzzper})--(\ref{xipar})].~In this
particular case, the Markovian approximation is applicable for
sure. The latter one is known to be an excellent approximation for
describing the radiative decay of an excited atom in free space
where the atom-field coupling strength is weak enough for atomic
motion memory effects to be insignificant, so that they may be
neglected~\cite{Heitler}. The time-dependent factor in the
kernel~(\ref{kerneldimless}) may then be replaced by its long-time
approximation (see, e.g.,~\cite{Davydov})
\begin{equation}
\frac{e^{-i(x-\tilde{x}_{A})(\tau-\tau{^\prime})}-1}{i(x-\tilde{x}_{A})}\rightarrow
-\pi\delta(x-\tilde{x}_{A})+i\frac{{\cal{P}}}{x-\tilde{x}_{A}}\,,
\label{Markov}
\end{equation}
(${\cal{P}}$ denotes a principal value) where
$\tilde{x}_{A}\!\approx\!x_{A}$ according to Eq.~(\ref{tildexA})
with small enough $\overline{\xi}^{\,\perp}(\mathbf{r}_{A},x)$. In
so doing, the kernel (\ref{kerneldimless}) becomes
\begin{equation}
K(\tau-\tau{^\prime})=-\frac{1}{2}\,\tilde{\Gamma}(x_{A})+i\Delta(x_{A})
\label{kernelMarkov}
\end{equation}
and the evolution equation (\ref{Volterradimless}) yields
\begin{equation}
C_{u}(\tau)=\exp\left[-\frac{1}{2}\,\tilde{\Gamma}(x_{A})+i\Delta(x_{A})\right]\tau
\label{Cumark}
\end{equation}
-- the exponential decay dynamics of the upper atomic level
shifted by
\begin{equation}
\Delta(x_{A})=\frac{{\cal{P}}}{2\pi}\!\int_{0}^{\infty}\!\!\!
dx\,\frac{\tilde{\Gamma}_{0}(x)}{x-x_{A}}\left[\left(\!\frac{x_{A}}{x_{}}\!\right)^{\!2}\!
\overline{\xi}^{\perp}(\mathbf{r}_{A},x)+\overline{\xi}^{\,\parallel}(\mathbf{r}_{A},x)\right]
\label{shift}
\end{equation}
[here, a~divergent contribution to the vacuum Lamb
shift has been omitted for the same reason as that in
Eq.~(\ref{tildexA})] with the rate
\begin{equation}
\tilde{\Gamma}(x_{A})=\tilde{\Gamma}_{0}(x_{A})\left[
1+\overline{\xi}^{\perp}(\mathbf{r}_{A},x_{A})+
\overline{\xi}^{\,\parallel}(\mathbf{r}_{A},x_{A})\right]
\label{rate}
\end{equation}
accounting for the presence of the nanotube.

\subsubsection{Single-Resonance Approximation of the Local Photonic DOS}

Another approximation that admits an analytical solution of the
evolution problem (\ref{Volterradimless}),~(\ref{kerneldimless})
is the single-resonance approximation of the local photonic DOS.
Suppose the functions $\overline{\xi}^{\,\perp}(\mathbf{r}_{A},x)$
and $\overline{\xi}^{\,\parallel}(\mathbf{r}_{A},x)$ have sharp
peaks at $x=x_{r}$ [the fact that they have the same positions of
extrema is obvious from Eqs.~(\ref{ksiper}) and (\ref{ksipar});
see also the footnote on page~\pageref{onthexiperpar}].~For all
$x$ in the vicinity of~$x_{r}$, their shape may then be roughly
approximated by the Lorentzian functions with the same
half-width-at-half-maximum $\delta x_{r}$ of the form
\begin{equation}
\overline{\xi}^{\,\perp(\parallel)}(\mathbf{r}_{A},x)\approx
\overline{\xi}^{\,\perp(\parallel)}(\mathbf{r}_{A},x_{r})
\frac{\delta x_{r}^{2}}{(x-x_{r})^{2}+\delta x_{r}^{2}}\,.
\label{Lorentian}
\end{equation}

The integrand in the kernel (\ref{kerneldimless}) is seen to be
basically contributed by $x\sim \tilde{x}_{A}$, thus making it
possible to approximate the factor in the square brackets as
follows
\[
1+\overline{\xi}^{\,\perp}(\mathbf{r}_{A},x)+
\left(\!\frac{x}{x_{A}}\!\right)^{\!2}\!
\overline{\xi}^{\,\parallel}(\mathbf{r}_{A},x)\approx
1+\overline{\xi}^{\,\perp}(\mathbf{r}_{A},x)+
\left(\!\frac{\tilde{x}_{A}}{x_{A}}\!\right)^{\!2}\!
\overline{\xi}^{\,\parallel}(\mathbf{r}_{A},x)
\]
with $\tilde{x}_A$ given in an explicit form by
Eq.~(\ref{tildexA}). Keeping this and the Lorentzian approximation
(\ref{Lorentian}) in mind, one calculates the kernel
(\ref{kerneldimless}) analytically to linear approximation in
$\delta x_{r}$ to obtain
\begin{equation}
K(\tau-\tau^{\prime})\approx\tilde{\Gamma}(x_{r})\,\frac{\delta
x_{r}}{2}\,\frac{\exp[-i(x_{r}-i\delta x_{r}-\tilde{x}_{A})
(\tau-\tau^{\prime})]-1}{i(x_{r}-i\delta x_{r}-\tilde{x}_{A})}\,,
\;\;\;\tau>\tau^{\prime}
\label{kernelapp1res}
\end{equation}
with $\tilde{\Gamma}(x_{r})$ given by Eq.~(\ref{rate}).
Substituting this into Eq.~(\ref{Volterradimless}) and making the
differentiation of both sides of the resulting equation over time,
followed by the change of the integration order and one more time
differentiation, one straightforwardly arrives at a second order
ordinary homogeneous differential equation of the form
\begin{equation}
\mbox{\it\"{C}}_{u}(\tau)+i(x_{r}-i\delta x_{r}-\tilde{x}_{A})
\mbox{\it\.{C}}_{u}(\tau)+(X/2)^{2}\,C_{u}(\tau)=0\,,\label{Custrong}
\end{equation}
where
\begin{equation}
X=\sqrt{2\delta x_{r}\tilde{\Gamma}(x_{r})}\,,\label{X}
\end{equation}
with the solution given for $\tilde{x}_{A}\approx x_{r}$ by
\begin{eqnarray}
C_{u}(\tau)\!\!\!&\approx&\!\!\!\frac{1}{2}\left(\!1+\frac{\delta
x_{r}} {\sqrt{\delta x_{r}^{2}-X^{2}}}\right)
\exp\!\left[-\!\left(\delta x_{r}-\sqrt{\delta x_{r}^{2}
-X^{2}}\right)\!\frac{\tau}{2}\,\right]\nonumber\\
&+&\!\!\!\frac{1}{2}\left(\!1-\frac{\delta x_{r}}{\sqrt{\delta
x_{r}^{2}-X^{2}}}\right)\exp\!\left[-\!\left(\delta
x_{r}+\sqrt{\delta
x_{r}^{2}-X^{2}}\right)\!\frac{\tau}{2}\,\right]. \label{Cuapp}
\end{eqnarray}
This solution is approximately valid for those atomic transition
frequencies $\tilde{x}_{A}$ which are located in the vicinity of
the photonic DOS resonances whatever the atom-field coupling
strength is.~In particular, if $(X/\delta x_{r})^{2}\!\ll\!1$,
Eq.~(\ref{Cuapp})~yields the exponential decay of the upper atomic
state population probability $|C_{u}(\tau)|^{2}$ with the rate
$\tilde{\Gamma}(x_{r})$ in full agreement with Eq.~(\ref{Cumark})
obtained within the Markovian approximation for weak atom-field
coupling.~In the opposite case, when $(X/\delta
x_{r})^{2}\!\gg\!1$, one has
\begin{equation}
|C_{u}(\tau)|^{2}\approx e^{-\delta x_{r}\tau}
\cos^{2}\!\left(\!\frac{X\tau}{2}\!\right),
\end{equation}
and the decay of the upper atomic state population probability
proceeds via damped Rabi oscillations with the dimensionless Rabi
frequency $X$ given by Eq.~(\ref{X}).~This is~the principal
signature of strong atom-field coupling which is beyond the
Markovian approximation.~Expressing $\tilde{\Gamma}(x_{r})$ in
Eq.~(\ref{X}) in terms of the local photonic DOS functions by
means of Eqs.~(\ref{rate}),~(\ref{xiper}),~(\ref{xipar}) and
approximating for atomic systems with Coulomb interaction
$\tilde{\Gamma}_{0}(x)\approx137^{-3}x$~\cite{Davydov}, one may
conveniently rewrite the strong atom-field coupling condition in
the form
\begin{equation}
2x_{r}\frac{\xi^{\perp}(x_{r})+\xi^{\parallel}(x_{r})}
{137^{3}\,\delta x_{r}}\gg1, \label{condstrong}
\end{equation}
from which it follows that the strong coupling regime is fostered
by high and narrow resonances in the local photonic DOS.

\subsection{Numerical Results and Discussion}\label{spontnumerically}

To get beyond the Markovian and single-resonance approximations,
we have solved Eqs.~(\ref{Volterradimless})
and~(\ref{kerneldimless}) numerically.~The \emph{exact} time
evolution of the upper state population probability
$|C_{u}(\tau)|^{2}$ was obtained for the atom placed [in a way
that Eq.~(\ref{condition}) was always satisfied] in the center and
near the wall inside, and at different distances outside achiral
CNs of different radii.~The local photonic DOS functions were
computed according to Eqs.~(\ref{xiper})--(\ref{ksipar}).~The CN
surface axial conductivity $\sigma_{zz}$ appearing in
Eqs.~(\ref{ksiper}),~(\ref{ksipar}) was calculated in the
relaxation-time approximation with the relaxation time
$3\times10^{-12}$~s; the spatial dispersion of $\sigma_{zz}$ was
neglected~\cite{Slepyan}.~The vacuum contribution to the kernel
(\ref{kerneldimless}) (that coming from the unity in the square
brackets) was taken into account within the Markovian
approximation (\ref{Markov}) with the divergent vacuum Lamb shift
omitted as the true (renormalized) vacuum Lamb shift may be
considered as included in the 'bare' atomic transition frequency
$x_{A}$~\cite{Dung,Welsch}. The Markovian approximation is known
to be an excellent approximation for describing atom-radiation
interaction processes in free space~\cite{Heitler}.~The vacuum
spontaneous decay rate was approximated by the expression
$\tilde{\Gamma}_{0}(x)\approx137^{-3}x$ valid for atomic systems
with Coulomb interaction~\cite{Davydov}.

%%%%%%%%%%%%%%%%%%%%%%%%%%%%%%%%%%%%%%%%%%%%%%%%%%%%%%%%%%%%%%%%%%%%%%%%%
\begin{figure}[p]
\begin{center}
\begin{minipage}[h]{120mm}
\epsfig{file=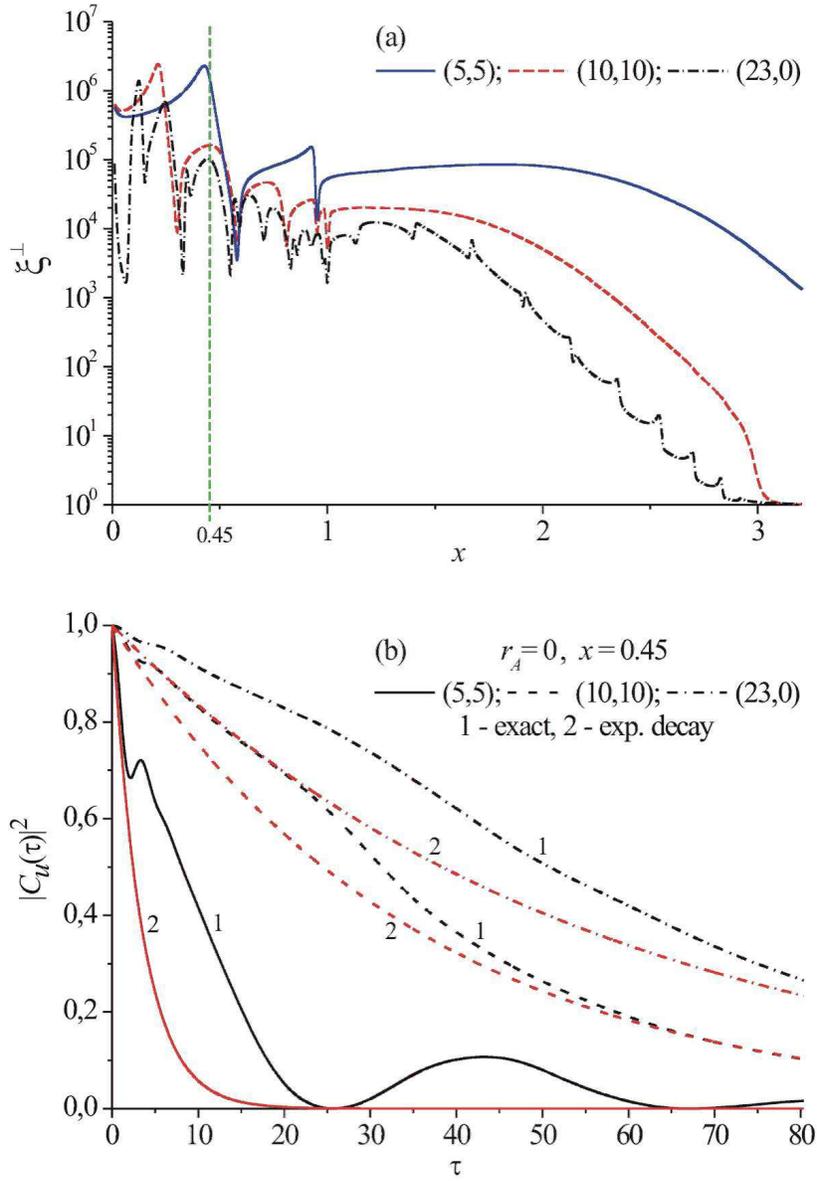, width=120mm}
\end{minipage}
\caption{Transverse local photonic DOS's~(a) and upper-level
spontaneous decay dynamics (b) for the atom in the center of
different CNs.~The ('bare') atomic transition frequency is
indicated by the dashed line in Fig.~\ref{fig3}~(a).} \label{fig3}
\end{center}
\end{figure}
%%%%%%%%%%%%%%%%%%%%%%%%%%%%%%%%%%%%%%%%%%%%%%%%%%%%%%%%%%%%%%%%%%%%%%%%%%

%%%%%%%%%%%%%%%%%%%%%%%%%%%%%%%%%%%%%%%%%%%%%%%%%%%%%%%%%%%%%%%%%%%%%%%%%
\begin{figure}[p]
\begin{center}
\begin{minipage}[h]{120mm}
\epsfig{file=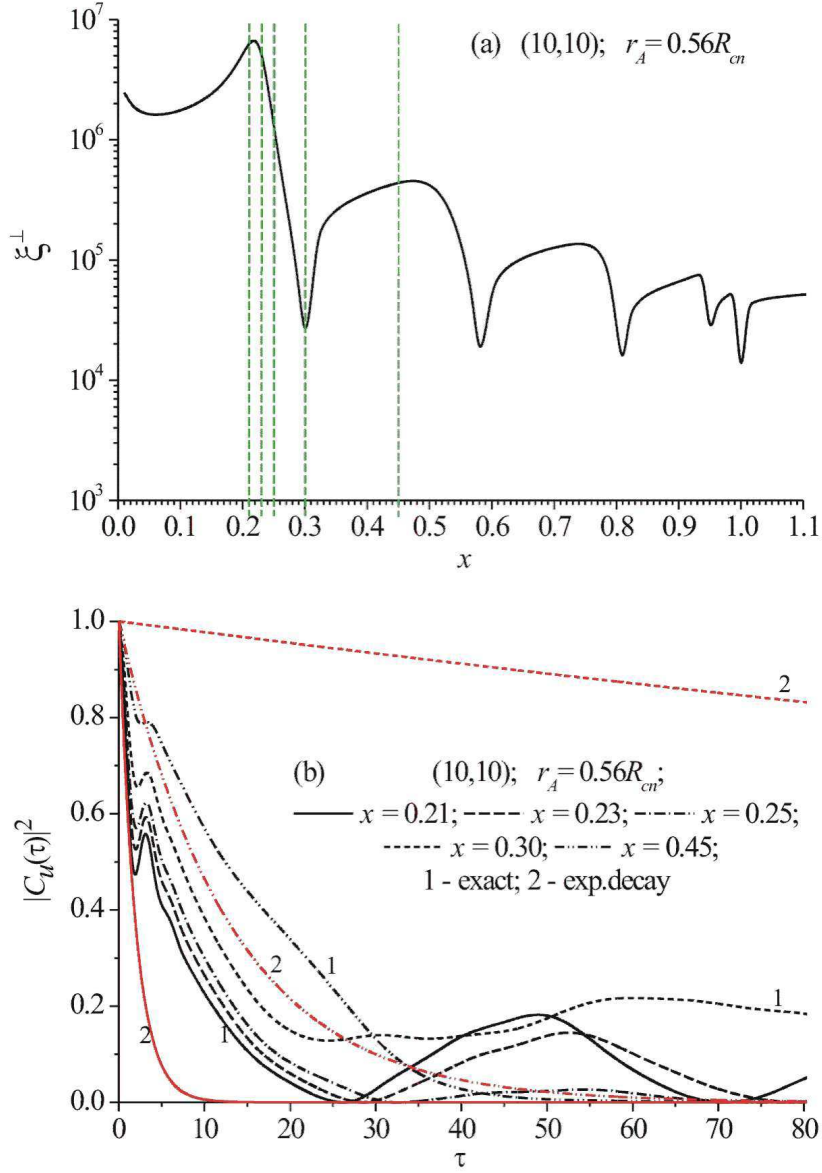, width=120mm}
\end{minipage}
\caption{(a)~Fragment of the transverse local photonic DOS for the
atom inside the (10,10) CN at distance of 3~\AA~from the wall (the
situation observed experimentally for Cs in
Ref.~\cite{Jeong}).~(b)~Upper-level spontaneous decay dynamics for
different ('bare') atomic transition frequencies [indicated by the
dashed lines in Fig.~\ref{fig4}~(a)] in this particular case.}
\label{fig4}
\end{center}
\end{figure}
%%%%%%%%%%%%%%%%%%%%%%%%%%%%%%%%%%%%%%%%%%%%%%%%%%%%%%%%%%%%%%%%%%%%%%%%%%

%%%%%%%%%%%%%%%%%%%%%%%%%%%%%%%%%%%%%%%%%%%%%%%%%%%%%%%%%%%%%%%%%%%%%%%%%
\begin{figure}[p]
\vskip-0.7cm
\begin{center}
\begin{minipage}[h]{80mm}
\epsfig{file=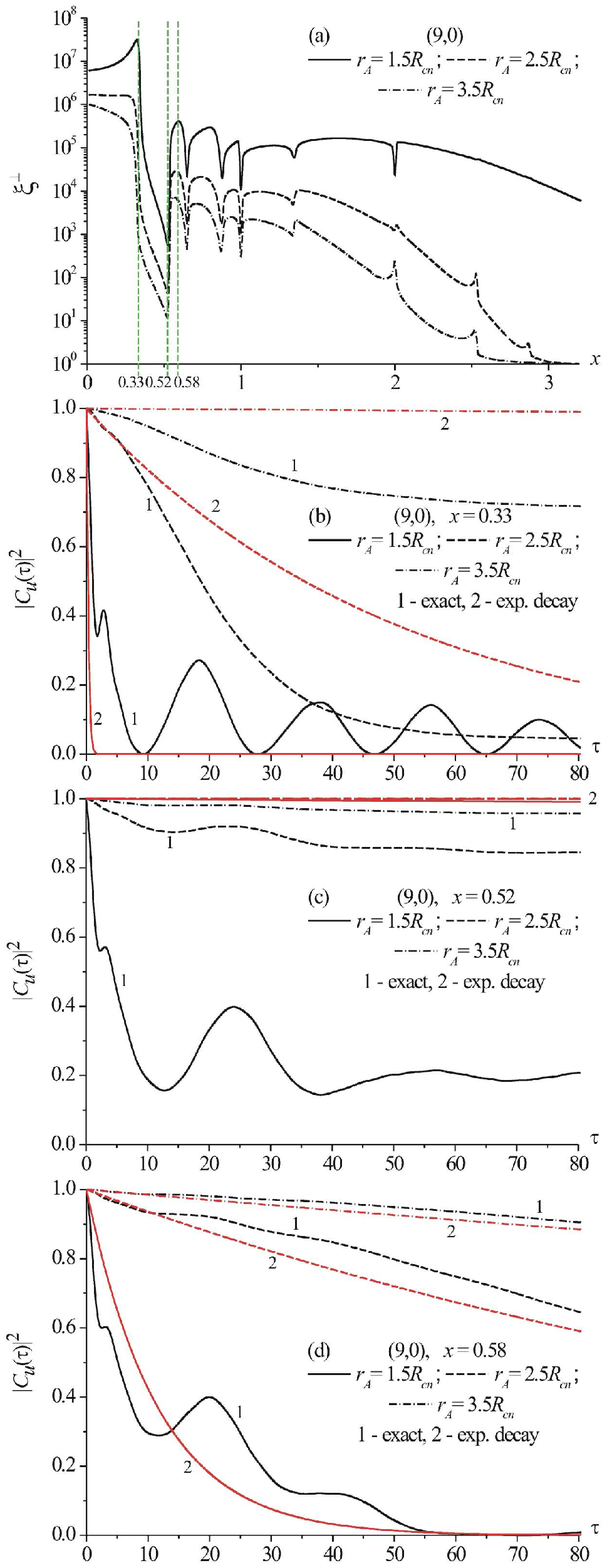, width=80mm}
\end{minipage}
\vskip-0.5cm \caption{(a)~Transverse local photonic DOS's for the
atom located at different distances outside the (9,0) CN. (b, c,
d)~Upper-level spontaneous decay dynamics for the three ('bare')
atomic transition frequencies [indicated by the dashed lines in
Fig.~\ref{fig5}~(a)] at different atom-nanotube-surface
distances.} \label{fig5}
\end{center}
\end{figure}
%%%%%%%%%%%%%%%%%%%%%%%%%%%%%%%%%%%%%%%%%%%%%%%%%%%%%%%%%%%%%%%%%%%%%%%%%%

Figure~\ref{fig3}~(a) presents $\xi^{\perp}(x)$ for the atom in
the center of the (5,5), (10,10) and (23,0) CNs.~It is seen to
decrease with increasing the CN radius, representing the decrease
of the atom-field coupling strength as the atom moves away from
the CN surface. To calculate $|C_{u}(\tau)|^{2}$ in this
particular case, we have fixed $x_{A}\!=\!0.45$ (indicated by the
vertical dashed line), firstly, because this transition is located
within the visible light range $0.305\!<\!x\!<\!0.574$, secondly,
because this is the approximate peak position of $\xi^{\perp}(x)$
for all the three CNs.~The functions $|C_{u}(\tau)|^{2}$
calculated are shown in Figure~\ref{fig2}~(b) in comparison with
those obtained in the Markovian approximation yielding the
exponential decay.~The actual spontaneous decay dynamics is
clearly seen to be non-exponential.~For the small radius (5,5) CN,
Rabi oscillations are observed, indicating a \emph{strong}
atom-field coupling regime related to \emph{strong} non-Markovian
memory effects.~Eq.~(\ref{condstrong}) is satisfied in this
case.~With increasing the CN radius, as the value of
$\xi^{\perp}(0.45)$ decreases, Eq.~(\ref{condstrong}) ceases to be
valid and the decay dynamics approaches the exponential one with
the decay rate enhanced by several orders of magnitude compared
with that in free space.~Note that, though the distance from the
atom to the CN surface is larger for the (23,0) CN than for the
(10,10) CN, the deviation of the actual decay dynamics from the
exponential one is larger for the (23,0) CN. This is an obvious
consequence of the influence of a~small neighboring peak in the
(23,0) CN photonic density of states [Figure~\ref{fig3}~(a)].

In Ref.~\cite{Jeong}, formation of Cs-encapsulating single-wall
CNs was reported.~In a~particular case of the~(10,10) CN, the
stable Cs atom/ion position was observed to be at distance of
3~\AA~from the wall.~We have simulated the spontaneous decay
dynamics for a number of typical atomic transition frequencies for
this case.~Figure~\ref{fig4}~(a) shows the local photonic DOS and
the five specific transition frequencies $x_{A}$ (dashed lines)
for which the functions $|C_{u}(t)|^{2}$ presented in
Figure~\ref{fig4}~(b) were calculated.~Rabi oscillations are
clearly seen to occur in the vicinity of the highest peak
($x_{A}\!\approx\!0.22$) of the photonic density of
states.~Important is that they persist for large enough detuning
values $x_{A}\!\approx\!0.21\!\div\!0.25$.~For $x_{A}\!=\!0.30$,
the density of photonic states has a dip, and the decay dynamics
exhibits no Rabi oscillations, being strongly non-exponential
nevertheless. For $x_{A}\!=\!0.45$, the intensity of the peak of
the photonic density of states is not large enough and the peak is
too broad, so that the strong atom-field coupling condition
(\ref{condstrong}) is not satisfied and the decay dynamics is
close to the exponential one.

Figure~\ref{fig5}~(a) shows the local photonic DOS for the atom
outside the (9,0) CN at the different distances from its
surface.~The vertical dashed lines indicate the atomic transitions
for which the functions $|C_{u}(\tau)|^{2}$ in
Figures~\ref{fig5}~(b),~(c), and (d) were calculated.~The
transitions $x_{A}\!=\!0.33$ and $0.58$ are the positions of sharp
peaks (at least for the shortest atom-surface distance), while
$x_{A}\!=\!0.52$ is the position of a dip of the function
$\xi^{\perp}(x)$.~Very clear underdamped Rabi oscillations are
seen for the shortest atom-surface distance at $x_{A}\!=\!0.33$
[Figure~\ref{fig4}~(b)], indicating strong atom-field coupling
with strong non-Markovity.~For $x_{A}\!=\!0.58$
[Figure~\ref{fig5}~(d)], the value of $\xi^{\perp}(0.58)$ is not
large enough for strong atom-field coupling to occur, so that
Eq.~(\ref{condstrong}) is not fulfilled.~As a consequence, the
decay dynamics, being strongly non-exponential in general,
exhibits no clear Rabi oscillations.~For $x_{A}\!=\!0.52$
[Figure~\ref{fig5}~(c)], though $\xi^{\perp}(0.52)$ is
comparatively small, the spontaneous decay dynamics is still
non-exponential, approaching the exponential one only when the
atom is far enough from the CN surface.

The reason for non-exponential spontaneous decay dynamics in all
the cases considered is similar to that taking place in photonic
crystals~\cite{John}.~When the atom is close enough to the CN
surface, an absolute value of the relative density of photonic
states is large and its frequency variation in the neighborhood of
a specific atomic transition frequency essentially influences the
time behavior of the kernel~(\ref{kerneldimless}) of the evolution
equation~(\ref{Volterradimless}).~Physically, this means that the
correlation time of the electromagnetic vacuum is not negligible
on the time scale of the evolution of the atomic system, so that
atomic motion memory effects are important and the Markovian
approximation in the kernel~(\ref{kerneldimless}) is
inapplicable.~The drastic increase of the local photonic DOS near
a CN can be interpreted as being due to the electromagnetic vacuum
renormalization: the relative density of photonic states near the
CN effectively increases since, along with ordinary free photons,
there appear surface photonic states coupled with CN electronic
quasiparticle excitations.~These latter ones are responsible for
the nonradiative atomic decay (photon emission by the atom with
subsequent CN quasiparticle
excitation)~\cite{Bondarev02}.~Obviously, such a decay does not
respond sensitively to the microscopic radiation-field structure
in a close vicinity of the CN surface.~This, along with
Eq.~(\ref{condition}), justifies rather small atom-surface
distances we have considered within the macroscopic Green-tensor
approach.

\section{van der Waals Energy of a Ground-State Atom near\\ a Carbon Nanotube}\label{vdwenergy}

\begin{figure}[t]
\begin{center}
\begin{minipage}[h]{100mm}
\epsfig{file=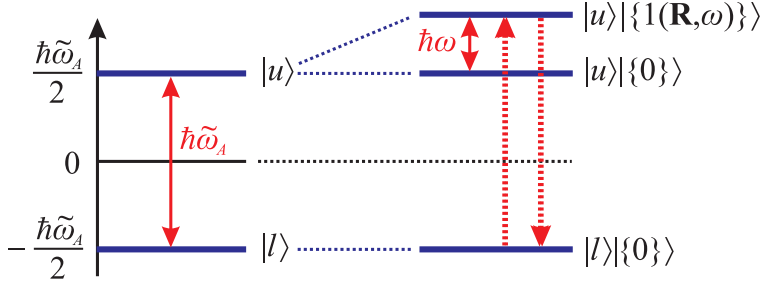, width=100mm}
\end{minipage}
\caption{Schematic of the energy levels in the problem of the
ground-state atom vdW interaction with the CN. On the left are the
unperturbed atomic levels given by the
Hamiltonian~(\ref{Harenorm}), on the right are the levels of the
coupled "atom--nanotube" system given by the total
Hamiltonian~(\ref{Htwolev})--(\ref{Hint}). The relevant virtual
transition of the system is shown by the vertical dashed arrows.
See Eqs.~(\ref{tildexA}) and (\ref{omegadimless})
for~$\tilde{\omega}_{A}$.} \label{fig6}
\end{center}
\end{figure}

\begin{figure}[t]
\begin{center}
\begin{minipage}[h]{120mm}
\epsfig{file=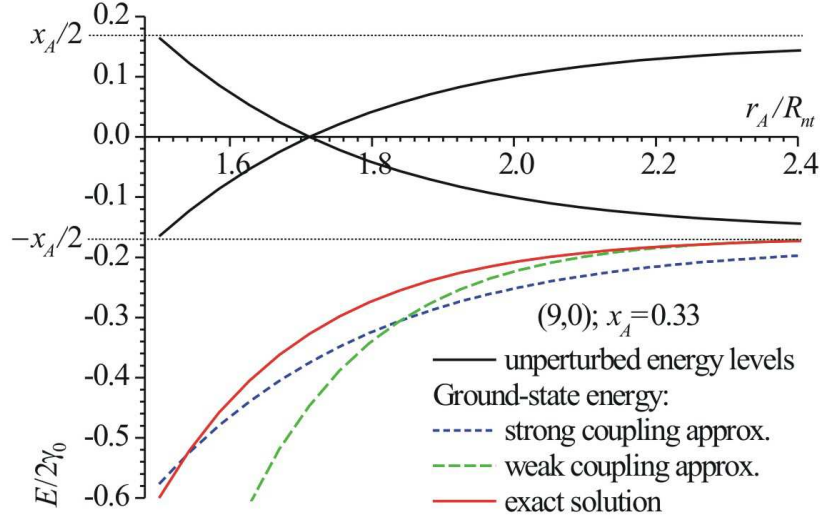, width=120mm}
\end{minipage}
\caption{Dimensionless unperturbed energy levels [given by the
Hamiltonian~(\ref{H0}) and (\ref{Harenorm})] and total
ground-state energy [given by Eqs.~(\ref{Evdw})--(\ref{Evw}),
(\ref{ksiper}) and (\ref{ksipar})] as functions of the atomic
position for the two-level atom outside the (9,0) CN. See
Eq.~(\ref{omegadimless}) for $x_{A}$.} \label{fig7}
\end{center}
\end{figure}

When the atom is in the ground state, its vdW energy near the
nanotube is given by the $\mathbf{r}_{A}$-dependent contribution
to the ground-state eigenvalue of the total
Hamiltonian~(\ref{Htwolev})-(\ref{Hint}).~Important, however, is
that the unperturbed atomic subsystem is now described by the
Hamiltonian (\ref{Harenorm}) with the renormalized transition
frequency~(\ref{omegarenorm}).~Eqs.~(\ref{tildexA}) and
(\ref{omegadimless}) represent the latter one in terms of the
transverse $\mathbf{r}_{A}$-dependent photonic DOS
$\overline{\xi}^{\,\perp}(\mathbf{r}_{A},x)$.~They indicate that
the transition frequency $\tilde{\omega}_{A}$ decreases with
increasing $\overline{\xi}^{\,\perp}$, i.~e.~when the atom
approaches the CN surface [see Fig.~\ref{fig5}~(a) as an example],
thereby bringing the unperturbed atomic levels together, or even
making them degenerated if $\overline{\xi}^{\,\perp}$ is large
enough [see the schematic of the energy levels of the problem in
Fig.~\ref{fig6} and a typical example of the (9,0) CN in
Fig.~\ref{fig7}].~If one uses the standard perturbation theory,
this would yield the divergence of high-order corrections to the
vdW energy at small atom-CN-surface distances because of the small
energy denominators in the corresponding power-series
expansions.~To account for a possible degeneracy of the
unperturbed atomic levels in a correct way, one has to use the
perturbation theory for degenerated atomic levels (see,
e.g.,~\cite{Davydov}).~In so doing, one has to take the upper
state degeneracy into account of the whole system with respect to
$\mathbf{R}$ as well as its possible degeneracy with respect to
$\omega$ when $\omega\sim\tilde{\omega}_{A}\sim0$ as is seen in
Fig.~\ref{fig6}. Thus, the ground-state wave function of the whole
system should be represented as a coherent mixture of the lower
and upper states of the form
\begin{equation}
|\psi\rangle=C_{l}\,|l\rangle|\{0\}\rangle+
\int_{0}^{\infty}\!\!\!\!\!d\omega\!\int\!d\mathbf{R}\,
C_{u}(\mathbf{R},\omega)\,|u\rangle|\{1(\mathbf{R},\omega)\}\rangle\,,
\label{wfunc}
\end{equation}
where $C_{l}$ and $C_{u}(\mathbf{R},\omega)$ are unknown mixing
coefficients of the lower and upper states of the \emph{whole}
system.

%%%%%%%%%%%%%%%%%%%%%%%%%%%%%%%%%%%%%%%%%%%%%%%%%%%%%%%%%%%%%%%%%%%%%%%%%
%%%%%%%%%%%%%%%%%%%%%%%%%%%%%%%%%%%%%%%%%%%%%%%%%%%%%%%%%%%%%%%%%%%%%%%%%%

%%%%%%%%%%%%%%%%%%%%%%%%%%%%%%%%%%%%%%%%%%%%%%%%%%%%%%%%%%%%%%%%%%%%%%%%%
%%%%%%%%%%%%%%%%%%%%%%%%%%%%%%%%%%%%%%%%%%%%%%%%%%%%%%%%%%%%%%%%%%%%%%%%%%

The total energy $E$ of the ground state is now given by the
solution of a secular equation obtained by applying the
Hamiltonian~(\ref{Htwolev})-(\ref{Hint}) to the wave
function~(\ref{wfunc}).~This yields the integral equation
\begin{equation}
E=-\frac{\hbar\tilde{\omega}_{A}}{2}-
\int_{0}^{\infty}\!\!\!\!\!d\omega\int\!d\mathbf{R}\,
\frac{\displaystyle|\mbox{g}^{(-)}(\mathbf{r}_{A},\mathbf{R},\omega)|^{2}}
{\displaystyle\hbar\omega+\frac{\hbar\tilde{\omega}_{A}}{2}-E}\,,
\label{E}
\end{equation}
which the ground-state-atom vdW energy $E_{vw}(\mathbf{r}_{A})$ is
determined from by means of an obvious relationship
\begin{equation}
E=-\frac{\hbar\omega_{A}}{2}+E_{vw}(\mathbf{r}_{A}) \label{Evdw}
\end{equation}
with the first term representing the 'bare' (non-inte\-racting)
two-level atom in the ground-state.~This, upon recalling
Eqs.~(\ref{DOSparper}),~(\ref{xiper}),~(\ref{xipar}) and
(\ref{omegadimless}) and introducing the dimensionless vdW energy
\begin{equation}
\varepsilon_{vw}(\mathbf{r}_{A})=\frac{E_{vw}(\mathbf{r}_{A})}{2\gamma_{0}}\,,
\label{energydimless}
\end{equation}
reduces to the integral equation of the form
\begin{eqnarray}
\varepsilon_{vw}(\mathbf{r}_{A})\!\!\!&=&\!\!\!\frac{x_{A}}{2\pi}
\int_{0}^{\infty}\!\!\!dx\,\frac{\tilde{\Gamma}_{0}(x)}{x^{2}}\,
\overline{\xi}^{\,\perp}(\mathbf{r}_{A},x)\label{Evw}\\
&-&\!\!\!\frac{x^{2}_{A}}{2\pi}\int_{0}^{\infty}\!\!\!dx\,\frac{\tilde{\Gamma}_{0}(x)}{x^{2}}\,
\frac{\displaystyle\overline{\xi}^{\,\perp}(\mathbf{r}_{A},x)+
{\left(\frac{x}{x_{A}}\right)^{\!2}}\overline{\xi}^{\,\parallel}(\mathbf{r}_{A},x)}
{\displaystyle x+x_{A}\left[1-\frac{1}{2\pi}
\int_{0}^{\infty}\!\!\!dx\,\frac{\tilde{\Gamma}_{0}(x)}{x^{2}}\,
\overline{\xi}^{\,\perp}(\mathbf{r}_{A},x)\right]
-\varepsilon_{vw}(\mathbf{r}_{A})}\,,\nonumber
\end{eqnarray}
representing the dimensionless ground-state-atom vdW energy in
terms of the transverse and longitudinal
$\mathbf{r}_{A}$-dependent (local) photonic DOS functions
$\overline{\xi}^{\,\perp}(\mathbf{r}_{A},x)$ and
$\overline{\xi}^{\,\parallel}(\mathbf{r}_{A},x)$, which are given
in an explicit form by Eqs.~(\ref{ksiper}) and (\ref{ksipar}),
respectively.

\subsection{Qualitative Analysis}\label{vdwqualitatively}

Eqs.~(\ref{Evdw})--(\ref{Evw}) along with Eqs.~(\ref{ksiper}) and
(\ref{ksipar}) describe the ground-state-atom vdW energy near an
infinitely long single-wall carbon nanotube in terms of the local
(distance-dependent) photonic DOS.~Eq.~(\ref{Evw}) is universal in
the sense that it covers both strong and weak atom-field coupling
regimes defined as those violating and non-violating,
respectively, the applicability domain of the conventional
stationary perturbation theory.\footnote{One should make a clear
difference between this definition and that where the strong
(weak) atom-field coupling regime is defined as that violating
(non-violating) the applicability domain of the Markovian
approximation~\cite{John}.~Our definition here is based on the
Born approximation associated with the strength of the coupling
between the atom and the photonic reservoir, whereas the Markovian
approximation is related to memory effects of the photonic
reservoir.}

\subsubsection{Weak Coupling Regime}

If, when the atom is faraway from the CN, the local photonic DOS
$\overline{\xi}^{\,\perp}(\mathbf{r}_{A},x)$ is such small that
\begin{equation}
\frac{1}{2\pi}\int_{0}^{\infty}\!\!\!dx\,\frac{\tilde{\Gamma}_{0}(x)}{x^{2}}\,
\overline{\xi}^{\,\perp}(\mathbf{r}_{A},x)\ll1 \label{DOSweakcond}
\end{equation}
and
\begin{equation}
|\varepsilon_{vw}(\mathbf{r}_{A})|\ll x_{A}\,, \label{vdwweakcond}
\end{equation}
then Eq.~(\ref{Evw}) yields a well-known perturbation theory
result (see, e.g.,~\cite{Buhmann04JOB,Buhmann04pre}) of the form
\begin{eqnarray}
\varepsilon_{vw}(\mathbf{r}_{A})\!\!\!&\approx&\!\!\!\frac{x_{A}}{2\pi}
\int_{0}^{\infty}\!\!\!dx\,\frac{\tilde{\Gamma}_{0}(x)}{x^{2}}\,
\overline{\xi}^{\,\perp}(\mathbf{r}_{A},x)\nonumber\\
&-&\!\!\!\frac{x^{2}_{A}}{2\pi}\int_{0}^{\infty}\!\!\!dx\,\frac{\tilde{\Gamma}_{0}(x)}{x^{2}}\,
\frac{\displaystyle\overline{\xi}^{\,\perp}(\mathbf{r}_{A},x)+
{\left(\frac{x}{x_{A}}\right)^{\!2}}\overline{\xi}^{\,\parallel}(\mathbf{r}_{A},x)}
{\displaystyle x+x_{A}}\,, \label{Evwweak}
\end{eqnarray}
where the first term comes from the unperturbed Hamiltonian
(\ref{H0}),~(\ref{Harenorm}) and the second one is the second
order correction due to the perturbation (\ref{Hint}).~This can be
equivalently rewritten in the form
\begin{equation}
\varepsilon_{vw}(\mathbf{r}_{A})\approx\frac{x_{A}}{2\pi}
\int_{0}^{\infty}\!\!\!dx\,\frac{\tilde{\Gamma}_{0}(x)}{x(x+x_{A})}
\left[\overline{\xi}^{\,\perp}(\mathbf{r}_{A},x)-
\frac{x}{x_{A}}\,\overline{\xi}^{\,\parallel}(\mathbf{r}_{A},x)\right],
\label{Evwweak1}
\end{equation}
from which, in view of Eqs.~(\ref{ksiper}),~(\ref{ksipar}) and
basic properties of the modified cylindric Bessel functions~(see,
e.g.~\cite{Abramovitz,Watson}), one can immediately come to an
interesting conclusion.~Namely, if the atom is fixed outside
(inside) the CN in a way that the weak atom-field coupling regime
is realized, i.e.~far enough from the CN surface, then the modulus
of the atom-nanotube vdW energy increases (decreases) with the CN
radius.~The conclusion is physically clear as the effective
atom-nanotube interaction area is larger (smaller) for
larger-radius nanotubes when the atom is outside (inside) the
CN.~Important, however, is that this obvious conclusion is,
strictly speaking, only valid in the weak atom-field coupling
regime for the outside atomic position, while for the inside
atomic position it represents a general effect of the effective
interaction area reduction with lowering the CN surface curvature.

In the large CN radius limit, Eq.~(\ref{Evwweak1}) along with
Eqs.~(\ref{ksiper}) and (\ref{ksipar}) can be shown to reproduce a
well-known Casimir-Polder result~\cite{Casimir} for an atom near
an infinitely conducting plane (see Appendix~C).

\subsubsection{Strong Coupling Regime}

When the atom is close enough to the CN surface, the local
photonic DOS $\overline{\xi}^{\,\perp}(\mathbf{r}_{A},x)$ is
large, so that one might expect the condition
\begin{equation}
\frac{1}{2\pi}\int_{0}^{\infty}\!\!\!dx\,\frac{\tilde{\Gamma}_{0}(x)}{x^{2}}\,
\overline{\xi}^{\,\perp}(\mathbf{r}_{A},x)\sim1
\label{DOSstrongcond}
\end{equation}
to hold true.~Under this condition, Eq.~(\ref{Evw}) reduces to an
integral equation of the form
\begin{eqnarray}
\varepsilon_{vw}(\mathbf{r}_{A})\!\!\!&\approx&\!\!\!\frac{x_{A}}{2\pi}
\int_{0}^{\infty}\!\!\!dx\,\frac{\tilde{\Gamma}_{0}(x)}{x^{2}}\,
\overline{\xi}^{\,\perp}(\mathbf{r}_{A},x)\nonumber\\
&-&\!\!\!\frac{x^{2}_{A}}{2\pi}\int_{0}^{\infty}\!\!\!dx\,\frac{\tilde{\Gamma}_{0}(x)}{x^{2}}\,
\frac{\displaystyle\overline{\xi}^{\,\perp}(\mathbf{r}_{A},x)+
{\left(\frac{x}{x_{A}}\right)^{\!2}}\overline{\xi}^{\,\parallel}(\mathbf{r}_{A},x)}
{\displaystyle x-\varepsilon_{vw}(\mathbf{r}_{A})}
\label{Evwstrong}
\end{eqnarray}
valid in the strong atom-field coupling regime.

Eq.~(\ref{DOSstrongcond}) corresponds to the \emph{inverted}
unperturbed atomic levels described by the Hamiltonian
(\ref{Harenorm}) with $\tilde{\omega}_{A}$ given by
Eqs.~(\ref{tildexA}) and (\ref{omegadimless}) (see Fig.~\ref{fig7}
as an example). The levels are degenerated when the left-hand side
of Eq.~(\ref{DOSstrongcond}) is exactly $0.5$. A~noteworthy point
is that if one used the weak coupling approximation of
Eq.~(\ref{Evw}) for the atom close to the CN surface where
Eq.~(\ref{DOSweakcond}) is not satisfied and
Eq.~(\ref{DOSstrongcond}) holds true instead, then the result
would be divergent at the lower limit of integration.~To avoid
this fact, one has to use either the strong coupling approximation
(\ref{Evwstrong}) or the exact equation~(\ref{Evw}) to calculate
the vdW energy of the atom close to the nanotube surface.

\subsection{Numerical Results and Discussion}\label{vdwnumerically}

Using Eqs.~(\ref{Evdw})--(\ref{Evw}) along with (\ref{ksiper}) and
(\ref{ksipar}), we have simulated the total ground state energy of
the whole system "two-level atom~+~CN-modified vacuum
electromagnetic field" and the vdW energy of the two-level atom
nearby metallic and semiconducting carbon nanotubes of different
radii.~The dimensionless free-space spontaneous decay rate in
Eq.~(\ref{Evw}) was approximated by the expression
$\tilde{\Gamma}_{0}(x)\approx137^{-3}x$ valid for atomic systems
with Coulomb interaction~\cite{Davydov}.

Figure~\ref{fig7} shows the dimensionless ground state energy
level of the whole system [given by the corresponding eigenvalue
of the total Hamiltonian~(\ref{Htwolev})--(\ref{Hint}) and
represented by Eqs.~(\ref{Evdw})--(\ref{Evw})] and two
dimensionless energy levels of the unperturbed
Hamiltonian~(\ref{H0}),~(\ref{Harenorm}) as functions of the
atomic position outside the (9,0) CN.~As the atom approaches the
CN surface, its unperturbed levels come together, then get
degenerated and even inverted at a very small atom-surface
distance.~In so doing, the weak coupling approximation for the
ground state energy [given by
Eqs.~(\ref{Evdw}),~(\ref{energydimless}),~(\ref{Evwweak1}) and
(\ref{ksiper}),~(\ref{ksipar})] diverges near the surface, whereas
the strong coupling approximation
[Eqs.~(\ref{Evdw}),~(\ref{energydimless}),~(\ref{Evwstrong}) and
(\ref{ksiper}),~(\ref{ksipar})] yields a finite result.~The exact
solution reproduces the weak coupling approximation at large and
the strong coupling approximation at short atom-surface distances,
respectively.~Close to the nanotube surface, the exact solution is
seen to be a little bit lower than that given by the strong
coupling approximation~(\ref{Evwstrong}).~This is because the
condition~(\ref{DOSstrongcond}) turns into the inequality in this
region with the left-hand side larger than unity that is not taken
into account by the equation~(\ref{Evwstrong}).

%%%%%%%%%%%%%%%%%%%%%%%%%%%%%%%%%%%%%%%%%%%%%%%%%%%%%%%%%%%%%%%%%%%%%%%%%
\begin{figure}[p]
\begin{center}
\begin{minipage}[h]{120mm}
\epsfig{file=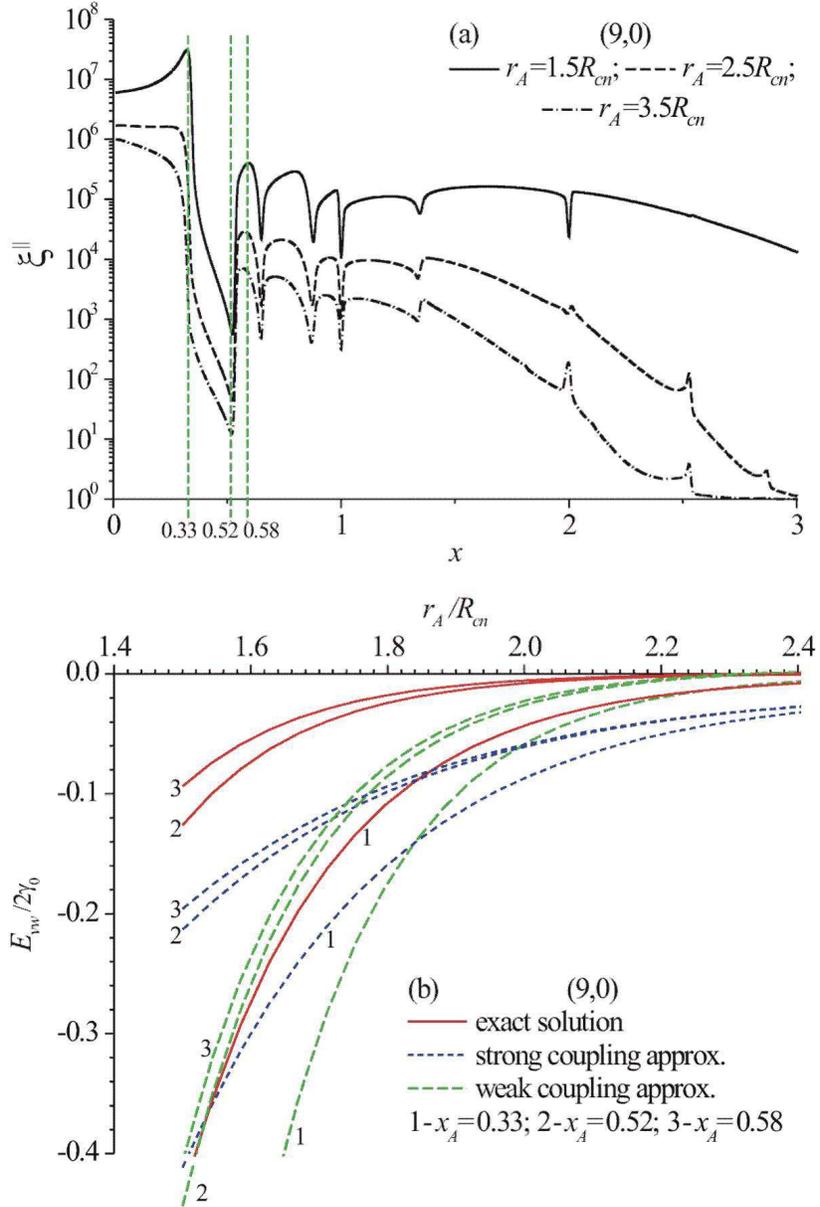, width=120mm}
\end{minipage}
\caption{Longitudinal local photonic DOS's (a) and ground-state
vdW energies~(b) as functions of the atomic position for the
two-level atom outside the (9,0) CN. The ('bare') atomic
transition frequencies are indicated by dashed lines in
Fig.~\ref{fig8}~(a).} \label{fig8}
\end{center}
\end{figure}
%%%%%%%%%%%%%%%%%%%%%%%%%%%%%%%%%%%%%%%%%%%%%%%%%%%%%%%%%%%%%%%%%%%%%%%%%%

%%%%%%%%%%%%%%%%%%%%%%%%%%%%%%%%%%%%%%%%%%%%%%%%%%%%%%%%%%%%%%%%%%%%%%%%%
\begin{figure}[p]
\begin{center}
\begin{minipage}[h]{120mm}
\epsfig{file=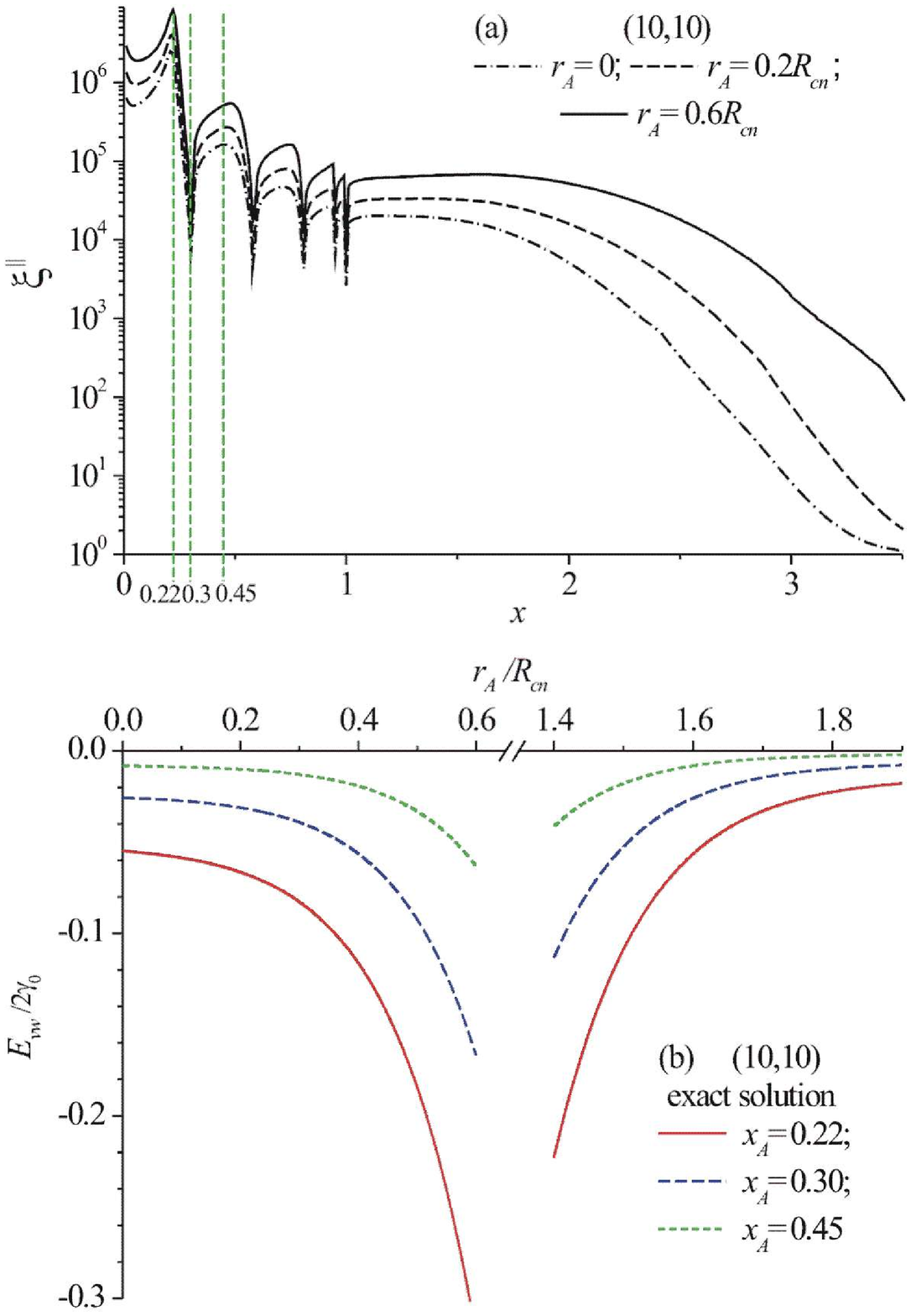, width=120mm}
\end{minipage}
\caption{Longitudinal local photonic DOS's (a) and ground-state
vdW energies~(b) as functions of the atomic position for the
two-level atom nearby the (10,10) CN. The ('bare') atomic
transition frequencies are indicated by dashed lines in
Fig.~\ref{fig9}~(a).} \label{fig9}
\end{center}
\end{figure}
%%%%%%%%%%%%%%%%%%%%%%%%%%%%%%%%%%%%%%%%%%%%%%%%%%%%%%%%%%%%%%%%%%%%%%%%%%

%%%%%%%%%%%%%%%%%%%%%%%%%%%%%%%%%%%%%%%%%%%%%%%%%%%%%%%%%%%%%%%%%%%%%%%%%
\begin{figure}[t]
\begin{center}
\begin{minipage}[h]{120mm}
\epsfig{file=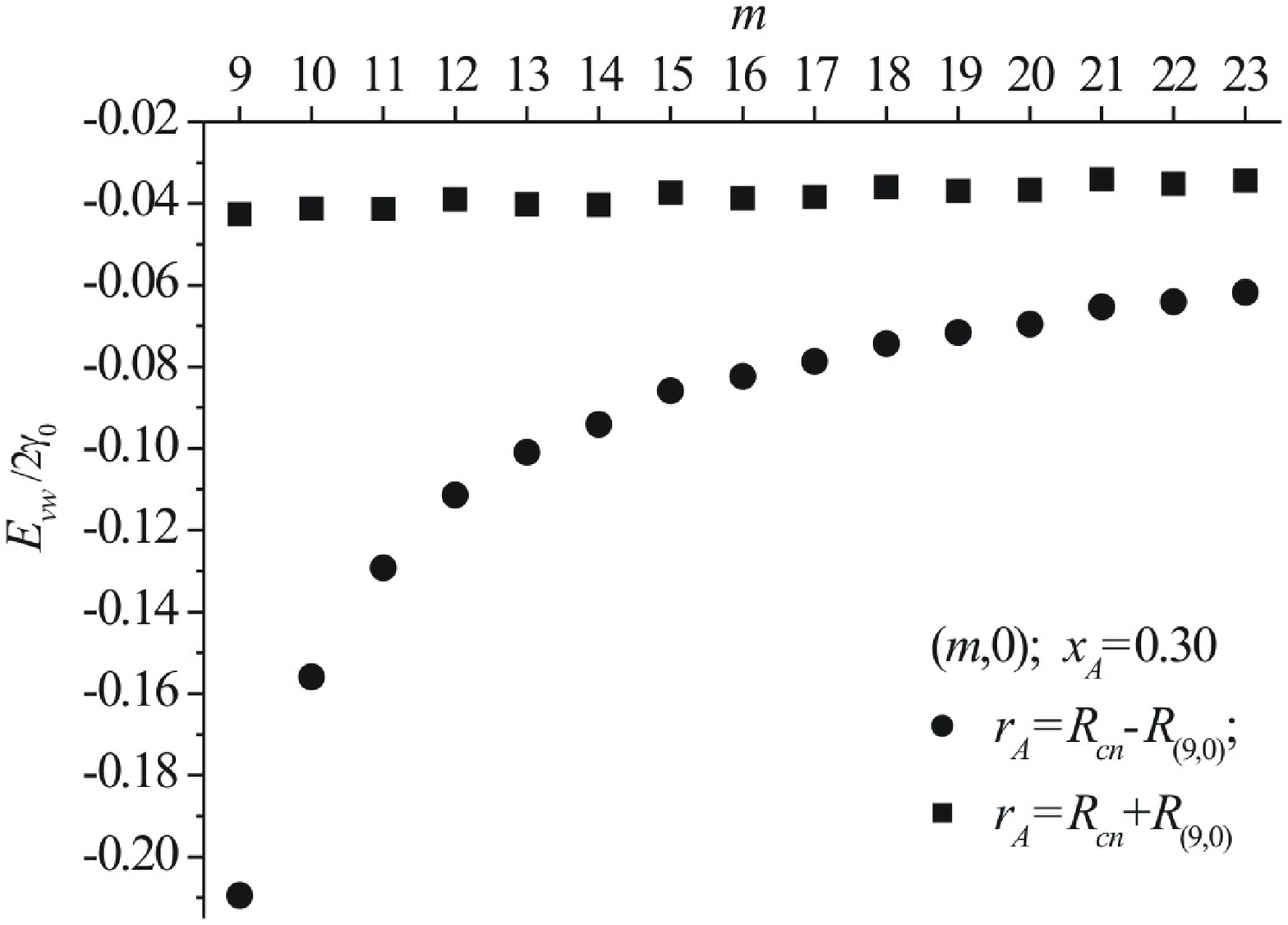, width=120mm}
\end{minipage}
\caption{Ground-state van der Waals energy of the two-level atom
positioned at a fixed atom-surface distance [equal to the radius
of the (9,0) CN] inside and outside "zigzag" $(m,0)$ CNs of
different radii, as a function of the nanotube index $m$.}
\label{fig10}
\end{center}
\end{figure}
%%%%%%%%%%%%%%%%%%%%%%%%%%%%%%%%%%%%%%%%%%%%%%%%%%%%%%%%%%%%%%%%%%%%%%%%%%

The degeneracy of the unperturbed atomic levels is the consequence
of the transverse local photonic DOS increase [see
Eq.~(\ref{tildexA})] as the atom approaches the CN surface.~A
typical example is shown in Fig.~\ref{fig5}~(a) for the atom
outside the (9,0) CN.~The transverse DOS function is seen to
increase with decreasing the atom-CN-surface distance,
representing the atom-field coupling strength enhancement with
surface photonic modes of the nanotube as the atom approaches the
nanotube surface.~The longitudinal local photonic DOS behaves in a
similar way as is seen in Fig.~\ref{fig8}~(a).\footnote{For the
small atom-CN-surface distances illustrated here, the
electromagnetic field retardation effects play no role and the
transverse and longitudinal DOS functions are actually exactly the
same. Indeed, taking the limit $c\!\rightarrow\!\infty$ allows one
to set $v=h$ in Eqs.~(\ref{xiperapp}) and (\ref{xiparapp}) in
Appendix~B, thereby making the transverse and longitudinal DOS
functions identical in the non-retarded limit. Both of these
functions are proportional to $c^3$, but this large factor is
cancelled out by the $c^3$ factor in the denominator of the
dimensionless vacuum spontaneous decay rate (\ref{G0x}) that
multiplies the photonic DOS functions in all the
expressions.\label{onthexiperpar}} The vertical dashed lines in
Fig.~\ref{fig8}~(a) indicate the 'bare' atomic transition
frequencies $x_{A}$ for which the vdW energies shown in
Fig.~\ref{fig8}~(b) were calculated.~Some of them are the
positions of peaks, the others are the positions of dips of the
local photonic DOS.~They are typical for some rear-earth ions such
as Europium, or for heavy hydrogen-like atoms such as Caesium
(supposed to be non-ionized near the CN). For
example~\cite{Schniepp}, for a~well-known two-level dipole
transition $^{7}F_{2}\leftrightarrow\,\!^{5}D_{0}$ of Europium at
$\lambda=615$~nm one has $x_{A}\approx0.37$, whereas for Caesium
one obtains from its first ionization potential~\cite{Lide} the
estimate
$x_{A}\approx3.89\mbox{\,eV}\times3/4\times(2\gamma_{0})^{-1}\approx0.5$
(the factor $3/4$ comes from the Lyman series of~Hydrogen), or
less for highly excited Rydberg states.

Figure~\ref{fig8}~(b) shows the vdW energies of the atom outside
the (9,0) CN for three different 'bare' atomic transition
frequencies indicated by vertical dashed lines in
Fig.~\ref{fig8}~(a).~Here, exact numerical solutions of
Eqs.~(\ref{Evdw})--(\ref{Evw}), (\ref{ksiper}), (\ref{ksipar}) are
compared with those obtained within the weak coupling
approximation (\ref{Evwweak1}),~(\ref{ksiper}),~(\ref{ksipar}) and
within the strong coupling approximation (\ref{Evwstrong}),
(\ref{ksiper}),~(\ref{ksipar}).~At small atom-CN-surface
distances, the exact solutions are seen to be fairly well
reproduced by those obtained in the strong coupling approximation,
clearly indicating the strong atom-field coupling regime in a
close vicinity of the nanotube surface.~The deviation from the
strong coupling approximation increases with transition
frequency~$x_{A}$, that is easily explicable since the degeneracy
condition (\ref{DOSstrongcond}) of the unperturbed atomic levels
is more difficult to reach for larger inter-level separations.~As
the atom moves away from the CN surface, the exact solutions
deviate from the strong coupling solutions and approach those
given by the weak coupling approximation, indicating the reduction
of the atom-field coupling strength with raising the atom-surface
distance.~The weak-coupling solutions are seen to be divergent
close to the nanotube surface as it should be because of the
degeneracy of the unperturbed atomic levels in this region.

Figure~\ref{fig9}~(a) shows the longitudinal local photonic DOS's
for the atom inside the (10,10) CN at three representative
distances from the nanotube wall.~As the atom approaches the wall,
the longitudinal DOS function increases in a similar way as that
Fig.~\ref{fig8}~(a) for the outside atomic position.~The vertical
dashed lines indicate the 'bare' atomic transition frequencies for
which the atomic vdW energies in Fig.~\ref{fig9}~(b) were
calculated.~Figure~\ref{fig9}~(b) represents the exact solutions
of Eq.~(\ref{Evw}) for atom inside and outside the (10,10)
CN.~When the atom is inside, the vdW energies are seen to be in
general lower than those for the atom outside the CN.~This may be
attributed to the fact that, due to the nanotube curvature, the
effective interaction area between the atom and the CN surface is
larger when the atom is inside rather than when it is outside the
CN.~This, in turn, indicates that encapsulation of doped atoms
into the nanotube is energetically more favorable than their
outside adsorption by the nanotube surface -- the effect observed
experimentally in Ref.~\cite{Jeong}.

Comparing Fig.~\ref{fig8}~(b) with Fig.~\ref{fig9}~(b) for the
outside atomic position, one can see that, if the atom-CN-surface
distance is so large that the weak-coupling approximation is good,
then for the same distance and approximately the same atomic
transition frequency $x_{A}\approx0.3$ the vdW energy near the
(10,10) nanotube ($R_{cn}=6.78$~\AA) is lower than that near the
(9,0) nanotube ($R_{cn}=3.52$~\AA).~This is in agreement with the
qualitative predictions formulated above in analyzing the weak
coupling regime.~In the present case, the effective
atom-CN-surface interaction area is larger for the atom outside
the larger-radius (10,10) nanotube, thereby explaining the
effect.~Obviously, one should expect an opposite effect for the
inside atomic position.~Experimentally, the heat of absorbtion of
Krypton, to take an example, has been shown to be larger for
graphite (0.17~eV/atom) than for multiwall nanotubes
(0.12~eV/atom)~\cite{Masenelli}, in agreement with our theoretical
predictions.

A similar tendency is demonstrated in Fig~\ref{fig10}.~Here, the
vdW energies calculated from Eqs.~(\ref{Evdw})--(\ref{Evw}),
(\ref{ksiper}), (\ref{ksipar}) are represented for the atom
positioned at a fixed distance from the nanotube wall [chosen to
be equal to the radius of the (9,0) CN] inside and outside the
"zigzag" $(m,0)$ CNs of increasing radius
$R_{cn}=m\,(a\sqrt{3}/2\pi)$.~When the atom is inside, the energy
absolute value goes down with $R_{cn}$, representing a general
tendency of the effective interaction area reduction with lowering
the CN surface curvature.~When the atom is outside, the effect
would be the opposite one if the atom were weakly coupled to the
field, thus indicating that the atom-field coupling regime is not
actually weak for the distance chosen.~In general, the inside and
outside vdW energies approach each other with $R_{cn}$ as it
should be since the two atomic positions become equivalent in the
plane limit where $R_{cn}\!\rightarrow\!\infty$.~Interesting is
also the fact that the vdW energies of the atom outside the
metallic nanotubes ($m$ is divisible by 3) are on average a little
bit smaller than those for the atom outside semiconducting
ones.~This is certainly the property related to the difference in
conductivities of metallic and semiconducting "zigzag"
nanotubes.~The property can be understood qualitatively as
follows.~In view of the absence of the longitudinal
depolarization~\cite{Benedict}, the CN dielectric tensor
$zz$-component $\epsilon_{zz}(\mathbf{R},\omega)$ is related to
the longitudinal polarizability $\alpha_{zz}(\omega)$ of the CN
per unit length via the equation
\[
\epsilon_{zz}(\mathbf{R},\omega)=1+4\pi\rho_{T}l_{T}\alpha_{zz}(\omega)
\]
with $l_{T}$ being the length of the tubule.~This, via the Drude
relationship~(\ref{sigmaCN}), yields
\[
\sigma_{zz}(\mathbf{R},\omega)=-i\omega\,\frac{\alpha_{zz}(\omega)}
{2\pi R_{cn}}\;
\]
from which it immediately follows that larger real conductivities
inherent to metallic CNs lead to larger imaginary polarizabilities
corresponding to stronger optical absorbtion by metallic CNs
compared with semiconducting ones.~It is then obvious that virtual
long-wavelength photon exchange responsible for the long-range
London-type dispersion vdW interaction is suppressed for the atom
outside the metallic CNs due to their stronger absorbtion compared
with the semiconducting CNs.

\section{Conclusion}\label{conclusion}

We have developed the quantum theory of near-field
electrodynamical properties of carbon nanotubes and investigated
spontaneous decay dynamics of excited states and van der Waals
attraction of the ground state of an atomic system (an atom or a
molecule) close to a~single-wall nanotube surface.~In describing
the atom-field interaction, we followed the electromagnetic field
quantization scheme developed for dispersing and absorbing media
in Refs.~\cite{Dung,Welsch}.~This quantization formalism was
adapted by us for a~particular case of an~atom near an infinitely
long single-wall~CN. We derived the simplified secondly quantized
Hamiltonian representing the "atom--nanotube" coupled system in
terms of only two standard approximations.~They are the electric
dipole approximation and the atomic two-level approximation.~The
(commonly used~\cite{Eberly}) rotating wave approximation was not
applied and the diamagnetic term of the atom-field interaction was
not neglected.~Starting with this general Hamiltonian, we have
obtained an evolution equation for the population probability of
the upper state and a vdW energy equation of the lower (ground)
state of the two-level atomic system coupled with the CN modified
vacuum electromagnetic field.~The equations are represented in
terms of the local photonic DOS and are valid for both strong and
weak atom-field coupling regime.

By solving the evolution equation of the upper state of the system
numerically, we have demonstrated a~strictly non-exponential
spontaneous decay dynamics in the case where the atom is close
enough to the CN surface.~In certain cases, namely when the atom
is close enough to the nanotube surface and the atomic transition
frequency is in the vicinity of the resonance of the local
photonic DOS, the system exhibits vacuum-field Rabi oscillations
-- a principal signature of strong atom-vacuum-field
coupling.~This is the result of strong non-Markovian memory
effects arising from the rapid frequency variation of the photonic
DOS near the nanotube.~The non-exponential decay dynamics gives
place to the exponential one if~the atom moves away from the CN
surface.~Thus, the atom-vacuum-field coupling strength and the
character of the spontaneous decay dynamics, respectively, may be
controlled~by changing the distance between the atom and CN
surface by means of a proper preparation of atomically doped CN
systems.

%%%%%%%%%%%%%%%%%%%%%%%%%%%%%%%%%%%%%%%%%%%%%%%%%%%%%%%%%%%%%%%%%%%%%%%%%
\begin{figure}[t]
\begin{center}
\begin{minipage}[h]{139mm}
\epsfig{file=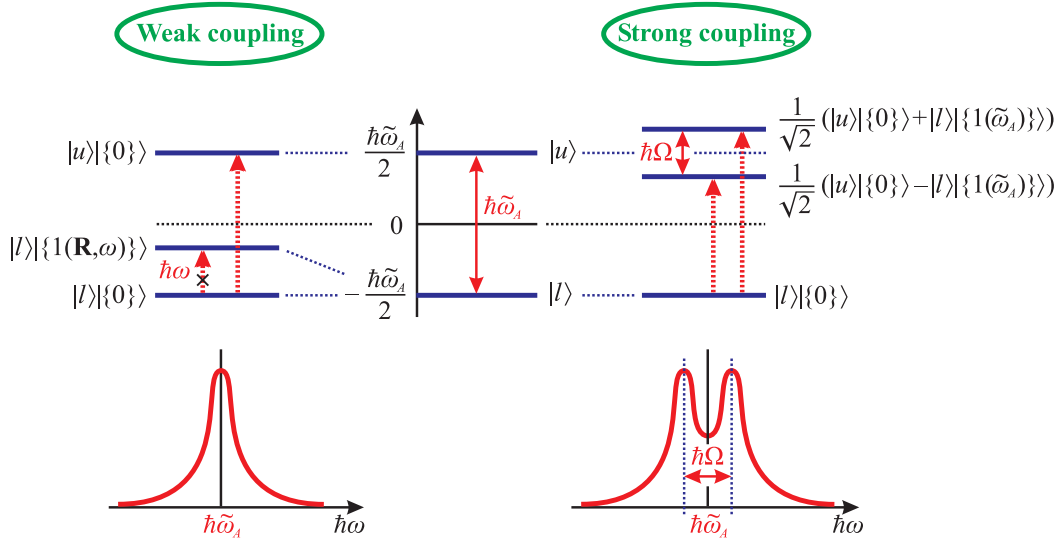, width=139mm}
\end{minipage}
\caption{Schematic of the energy levels and the absorbtion
line-shapes expected in the optical absorbance experiment with
atomically doped CNs.~In the center are the unperturbed atomic
levels given by the Hamiltonian~(\ref{Harenorm}).~On the left and
on the right are the levels of the coupled "atom--nanotube" system
in the weak and strong atom-vacuum-field coupling regime,
respectively.~The system is excited by an external optical
radiation. The allowed and forbidden (crossed) optical transitions
are shown by vertical dashed arrows. See Eqs.~(\ref{tildexA}),
(\ref{X}) and (\ref{omegadimless}) for $\tilde{\omega}_{A}$
and~$\Omega$.} \label{fig11}
\end{center}
\end{figure}
%%%%%%%%%%%%%%%%%%%%%%%%%%%%%%%%%%%%%%%%%%%%%%%%%%%%%%%%%%%%%%%%%%%%%%%%%%

We would like to emphasize a general character of the conclusion
above and its applied and fundamental significance.~We have shown
that similar to semiconductor
microcavities~\cite{Reithmaier,Peter,Sorba} and photonic band-gap
materials~\cite{John,Yoshie}, carbon nanotubes may qualitatively
change the character of the atom-electromagnetic-field
interaction, yielding strong atom-field coupling.~The study of
such phenomena was started awhile ago in atomic
physics~\cite{Haroche1} and still attracts a lot of interest in
connection with various quantum optics and nanophotonics
applications~\cite{Vuchkovich,Finley,Yamamoto,Noda} as well as
quantum computation and quantum information
processing~\cite{Weisbuch,Raimond,Abstreiter,Steel}. The fact that
the carbon nanotube may control atom-electromagnetic-field
coupling opens routes for new challenging applications of
atomically doped CN systems as various sources of coherent light
emitted by dopant atoms.

Strong atom-vacuum-field coupling we predict will yield an
additional structure in optical absorbance/reflectance spectra
(see, e.g.,~\cite{Li,Marinop}) of atomically doped CNs in the
vicinity of the energy of an atomic transition. The illustration
is shown in Fig.~\ref{fig11}. Weak non-Markovity of the
atom-vacuum-field interactions (that yielding non-exponential
spontaneous decay dynamics with no Rabi oscillations) will cause
an asymmetry of an optical spectral line-shape (not shown in
Fig.~\ref{fig11}) similar to that taking place for the exciton
optical absorbtion line-shape in quantum
dots~\cite{Bondarev03}.~Strong non-Markovity of the
atom-vacuum-field interactions (yielding non-exponential
spontaneous decay dynamics with fast Rabi oscillations) originates
from strong atom-vacuum-field coupling with the upper state of the
system splitted into two "dressed" states (see Fig.~\ref{fig11},
on the right).~This will yield a two-component structure of
optical absorbance/reflectance spectra similar to that observed
for excitonic and intersubband electronic transitions in
semiconductor quantum microcavities~\cite{Weisbuch1,Sorba}.

The atomic state strongly coupled to the surface vacuum photonic
modes of the nanotube is nothing but a 'quasi-1D cavity polariton'
similar by its physical nature to quasi-2D
excitonic~\cite{Weisbuch1}, intersubband electronic~\cite{Sorba}
and quasi-0D excitonic~\cite{Reithmaier,Peter} polaritons in
semiconductor quantum microcavities as well as to quasi-0D
excitonic polaritons in photonic crystal
nanocavities~\cite{Yoshie} recently observed experimentally.
Obviously, the stability of quasi-1D atomic polaritons in CNs is
basically determined by the atom-nanotube van der Waals
interaction.~This interaction, however, originates from
\emph{strong} atom-field coupling and, therefore, cannot be
correctly described in terms of vacuum-QED-based (weak-coupling)
vdW interaction models as well as in terms of those based upon the
linear response theory.~We have developed a simple quantum
mechanical approach to the ground-state vdW energy calculation of
a two-level atomic system near a CN.~The approach is based upon
the perturbation theory for degenerated atomic levels, thus
accounting for both weak and strong atom-field coupling and
thereby covering the vdW interactions of the ground-state quasi-1D
atomic polaritons in CNs.~Within the framework of this approach,
the vdW energy is described by the integral equation represented
in terms of the local photonic DOS.~By solving this equation
numerically, we have demonstrated the inapplicability of
weak-coupling-based vdW interaction models in a close vicinity of
the nanotube surface where the local photonic DOS effectively
increases, giving rise to an atom-field coupling enhancement
followed by the degeneracy of the unperturbed atomic levels due to
the diamagnetic interaction term.

The fundamental conclusion above is supplemented by those
important for various applications of atomically doped CNs in
modern nanotechnology.~In particular, we have studied CN surface
curvature effects on the atom-nanotube vdW interactions and have
shown that an inside encapsulation of doped atoms into the
nanotube is energetically more favorable than their outside
adsorption by the nanotube surface, in agreement with the
experimental observations reported in Ref.~\cite{Jeong}.~Moreover,
if the atom is fixed outside the CN in a way that the weak
atom-field coupling regime is realized, i.e.~far enough from the
CN surface, then the modulus of the atom-nanotube vdW energy
increases with the CN radius because the effective atom-nanotube
interaction area is larger for larger-radius nanotubes in this
case.~For inside atomic position, the modulus of the atom-nanotube
vdW energy decreases with the CN radius, representing a general
effect of the effective interaction area reduction with lowering
the CN surface curvature.

Finally, we here only dealt with the simplest manifestations of
the strong coupling regime in the atom-electromagnetic-field
interactions near carbon nanotubes. Similar manifestations of
strong atom-field coupling may occur in many other
atom-electromagnetic-field interaction processes in the presence
of CNs, such as, e.g., atomic states entanglement~\cite{Scheel},
interatomic dipole-dipole interactions~\cite{Agarwal,Dung02},
cascade spontaneous transitions in three-level atomic
systems~\cite{Dalton}, etc.~A further intriguing extension of the
present work could also be the study of the vdW interactions of
excited atomic states where, even in the weak atom-field-coupling
regime and in the simplest case of an atom near a planar
semi-infinite medium, very interesting peculiarities (e.g.,
oscillatory behavior) were recently shown to
exist~\cite{Buhmann04pre}.

\begin{center}
{\bf Acknowledgements}
\end{center}
The authors would like to thank G.~Abstreiter, I.D.~Feranchuk,
J.~Finley, G.Ya.~Sle\-pyan and D.-G.~Welsch for fruitful
discussions.

\appendix
\section{Two-level approximation}\label{2levelappr}

In this Appendix, we rewrite the
Hamiltonian~(\ref{Htot})--(\ref{Haf2}) in terms of a two-level
atomic model~\cite{Eberly}.~Within the framework of this model,
the spectrum of the atomic Hamiltonian~(\ref{Ha}) is approximated
by the two eigenstates, upper $|u\rangle$ and lower $|l\rangle$,
with the energies $\hbar\omega_{u}$ and $\hbar\omega_{l}$,
respectively.~One has then
\begin{equation}
\hat{H}_{A}\approx\hbar\omega_{u}|u\rangle\langle u|+
\hbar\omega_{l}|l\rangle\langle l| \label{Haapp1}
\end{equation}
with a completeness relation
\begin{equation}
|u\rangle\langle u|+|l\rangle\langle l|=\hat{I}.
\label{completeness}
\end{equation}
Subtracting a constant term
$(\hbar/2)(\omega_{u}+\omega_{l})\hat{I}$ from the
Hamiltonian~(\ref{Haapp1}) and thereby placing the energy zero in
the middle between the two energy levels, one arrives at the
'bare' two-level atomic Hamiltonian of the form
\begin{equation}
\hat{H}_{A}=\frac{\hbar\omega_{A}}{2}\,\hat{\sigma}_{z}
\label{Hbare}
\end{equation}
with $\hat{\sigma}_{z}=|u\rangle\langle u|-|l\rangle\langle l|$
and $\omega_{A}=\omega_{u}-\omega_{l}$ being the 'bare' atomic
transition frequency.

In terms of such a~two-level scheme, the atomic dipole moment
operator $\hat{\mathbf{d}}$ has the matrix elements $\langle
u|\hat{\mathbf{d}}|l\rangle=\langle
l|\hat{\mathbf{d}}|u\rangle=\mathbf{d}$ and $\langle
u|\hat{\mathbf{d}}|u\rangle=\langle
l|\hat{\mathbf{d}}|l\rangle=0$.~Keeping this and the completeness
relation~(\ref{completeness}) in mind, the interaction
Hamiltonians~(\ref{Haf1}) and (\ref{Haf2}) can be rewritten as
follows.

For the interaction~(\ref{Haf1}), using Eqs.~(\ref{vecpot}) and
(\ref{scalpot}) and a~well-known quantum mechanical operator
equality
\begin{equation}
(\hat{\mathbf{p}}_{i})_{\alpha}=m_{i}\frac{d}{dt}(\hat{\mathbf{r}}_{i})_{\alpha}
=\frac{m_{i}}{i\hbar}\,[(\hat{\mathbf{r}}_{i})_{\alpha},\hat{H}_{A}]\,,\\[0.25cm]
\label{p}
\end{equation}
where the Latin and Greek indexes enumerate particles in the atom
and vector components, respectively, and $\hat{H}_{A}$ is given by
Eq.~(\ref{Hbare}), one has
\begin{eqnarray}
\hat{H}^{(1)}_{AF}\!\!\!&=&\!\!\!(\hat{\sigma}-\hat{\sigma}^{\dag})\,\mathbf{d}\cdot\!\left[
\int_{0}^{\infty}\!\!\!\!\!d\omega\frac{\omega_{A}}{\omega}\,
\underline{\hat{\mathbf{E}}}^{\perp}(\mathbf{r}_{A},\omega)-\mbox{h.c.}\right]\nonumber\\
&-&\!\!\!(\hat{\sigma}+\hat{\sigma}^{\dag})\,\mathbf{d}\cdot\!\left[
\int_{0}^{\infty}\!\!\!\!\!d\omega\,
\underline{\hat{\mathbf{E}}}^{\parallel}(\mathbf{r}_{A},\omega)+\mbox{h.c.}\right],
\label{Haf1app}
\end{eqnarray}
with $\hat{\sigma}=|l\rangle\langle u|$ and
$\hat{\sigma}^{\dag}=|u\rangle\langle l|$.~Using further
Eqs.~(\ref{elperpar}), (\ref{Erw}) and (\ref{currentCN}) and
assuming the longitudinal (along the CN axis) atomic dipole
orientation due to the dominant axial polarizability of the
nanotube~\cite{Benedict,Tasaki,Jorio,Li,Marinop}, one arrives at
the secondly quantized interaction Hamiltonian~(\ref{Hint}) with
the interaction matrix elements (\ref{gpm}) and (\ref{gperppar}).

For the interaction~(\ref{Haf2}), one may proceed as follows
\begin{eqnarray}
\hat{H}^{(2)}_{AF}\!\!\!&=&\!\!\!\sum_{i,j}\sum_{\alpha,\beta}\frac{q_{i}q_{j}}{2m_{i}c^{2}}
\,\delta_{ij}\delta_{\alpha\beta}
\hat{A}_{\alpha}(\mathbf{r}_{A})\hat{A}_{\beta}(\mathbf{r}_{A})\nonumber\\
&=&\!\!\!\frac{i}{2\hbar
c^{2}}\sum_{i,\alpha,\beta}\frac{q_{i}}{m_{i}}
[(\hat{\mathbf{p}}_{i})_{\alpha},\hat{d}_{\beta}]
\hat{A}_{\alpha}(\mathbf{r}_{A})\hat{A}_{\beta}(\mathbf{r}_{A})\,,
\label{Haf2app1}
\end{eqnarray}
where the product of the two Kronecker-symbols in the first line
is represented in terms of the 'coordinate-momentum' commutator as
$$\delta_{ij}\delta_{\alpha\beta}=\frac{i}{\hbar}\,
[(\hat{\mathbf{p}}_{i})_{\alpha},(\hat{\mathbf{r}}_{j})_{\beta}].$$
Using Eq.~(\ref{p}) for the momentum operator components and the
completeness relation~(\ref{completeness}), one arrives at
\begin{eqnarray}
\hat{H}^{(2)}_{AF}=-\frac{\omega_{A}}{\hbar c^{2}}\,
\hat{\sigma}_{z}\!\left[\mathbf{d}\cdot
\hat{\mathbf{A}}(\mathbf{r}_{A})\right]^{2} \label{Haf2app2}
\end{eqnarray}
which, upon substituting Eq.~(\ref{vecpot}) for the vector
potential operator, becomes
\begin{eqnarray}
-\frac{\hat{\sigma}_{z}}{\hbar\omega_{A}}\sum_{\alpha,\beta}d_{\alpha}d_{\beta}
\left\{\int_{0}^{\infty}\!\!\!\!\!d\omega\frac{\omega_{A}}{i\omega}\!
\int_{0}^{\infty}\!\!\!\!\!d\omega^{\prime}\frac{\omega_{A}}{i\omega^{\prime}}\,
\underline{\hat{E}\!}_{\,\alpha}^{\,\perp}(\mathbf{r}_{A},\omega)\,
\underline{\hat{E}\!}_{\,\beta}^{\,\perp}(\mathbf{r}_{A},\omega^{\prime})\right.\nonumber\\
\left.-\!\int_{0}^{\infty}\!\!\!\!\!d\omega\frac{\omega_{A}}{i\omega}\!
\int_{0}^{\infty}\!\!\!\!\!d\omega^{\prime}\frac{\omega_{A}}{i\omega^{\prime}}\,
\underline{\hat{E}\!}_{\,\alpha}^{\,\perp}(\mathbf{r}_{A},\omega)\!
\left[\underline{\hat{E}\!}_{\,\beta}^{\,\perp}(\mathbf{r}_{A},\omega^{\prime})\right]^{\dag}
\!+\mbox{h.c.}\right\}.\nonumber
\end{eqnarray}
The four items here can, in view of Eqs.~(\ref{elperpar}),
(\ref{Erw}) and (\ref{currentCN}), be classified as follows.~The
first item and its hermitian conjugate describe the processes with
simultaneous annihilation and creation of \emph{two} photons
(two-photon transitions).~They may be safely neglected if the
field intensity is low enough that, we believe, is the case for
the vacuum electromagnetic field we deal with in considering the
vdW interactions.~Furthermore, keeping them in the Hamiltonian
$\hat{H}^{(2)}_{AF}$ is meaningless as the Hamiltonian
$\hat{H}^{(1)}_{AF}$ is nothing but a dipole approximation which,
by its definition, neglects two-photon transitions.~The second
item and its hermitian conjugate are combined together by using
bosonic commutation relations (\ref{commut}) to give in terms of
the bosonic field operators and the transverse dipole interaction
matrix element~(\ref{gperppar}) the expression
\begin{eqnarray}
-\frac{\hat{\sigma}_{z}}{\hbar\omega_{A}}\int_{0}^{\infty}\!\!\!\!\!d\omega\!
\int\!d\mathbf{R}\,|\mbox{g}^{\perp}(\mathbf{r}_{A},\mathbf{R},\omega)|^{2}\hskip3.5cm\nonumber\\
-\frac{2\hat{\sigma}_{z}}{\hbar\omega_{A}}\int_{0}^{\infty}\!\!\!\!\!d\omega\,
d\omega^{\prime}\!\int\!d\mathbf{R}\,d\mathbf{R}^{\prime}
\left[\mbox{g}^{\perp}(\mathbf{r}_{A},\mathbf{R},\omega)\right]^{\ast}\!
\mbox{g}^{\perp}(\mathbf{r}_{A},\mathbf{R}^{\prime},\omega^{\prime})
f^{\dag}(\mathbf{R},\omega)f(\mathbf{R}^{\prime},\omega^{\prime}),\nonumber
\end{eqnarray}
where the second term is nothing but a two-photon correction to
the dipole interaction~(\ref{Hint}), that can be easily seen by
noting that
$\hat{\sigma}_{z}=\hat{\sigma}^{\dag}\hat{\sigma}-\hat{\sigma}\hat{\sigma}^{\dag}$.~This
correction must be neglected for the reasons just discussed, so
that finally one arrives at
\begin{eqnarray}
\hat{H}^{(2)}_{AF}\approx-\frac{\hat{\sigma}_{z}}{\hbar\omega_{A}}
\int_{0}^{\infty}\!\!\!\!\!d\omega\!\int\!d\mathbf{R}\,
|\mbox{g}^{\perp}(\mathbf{r}_{A},\mathbf{R},\omega)|^{2}.
\label{Haf2app3}
\end{eqnarray}
Eq.~(\ref{Haf2app3}) can now be combined with the 'bare' atomic
Hamiltonian~(\ref{Hbare}) to yield the 'effective' unperturbed
atomic Hamiltonian~(\ref{Harenorm}) with the renormalized atomic
transition frequency given by Eq.~(\ref{omegarenorm}).

Thus, we have proved that the
Hamiltonian~(\ref{Htot})--(\ref{Haf2}), being approximated in
terms of the two-level atomic model, is represented by the
Hamiltonian~(\ref{Htwolev})--(\ref{Hint}).

\section{Green tensor and local photonic DOS}\label{transvlongDOS}

\subsubsection*{(a) Green tensor of a single-wall carbon nanotube}

We start with Eq.~(\ref{GreenequCN}) for the Green tensor of the
electromagnetic subsystem.~This equation is a direct consequence
of the equation
\begin{equation}
\sum_{\alpha=r,\varphi,z}\!\!\!
\left(\bm{\nabla}\!\times\bm{\nabla}\!\times-\,k^{2}\right)_{\!z\alpha}
\underline{\hat{E}\!}_{\,\alpha}(\mathbf{r},\omega)=
i\frac{4\pi}{c}\,k\,\underline{\hat{I}\!}_{\,z}(\mathbf{r},\omega)
\label{EIapp}
\end{equation}
obtained by substituting the magnetic field operator from
Eq.~(\ref{MaxwelE}) into Eq.~(\ref{MaxwelH}). On the other hand,
under the condition of the Coulomb gauge
(see,~e.g.,~\cite{Jackson}), one has
\begin{equation}
\underline{\hat{\mathbf{E}}}(\mathbf{r},\omega)=
\underline{\hat{\mathbf{E}}}^{\perp}(\mathbf{r},\omega)+
\underline{\hat{\mathbf{E}}}^{\parallel}(\mathbf{r},\omega)
\label{EAPhiapp}
\end{equation}
with [compare with Eqs.~(\ref{vecpot}) and (\ref{scalpot})]
\begin{equation}
\underline{\hat{\mathbf{E}}}^{\perp}(\mathbf{r},\omega)=
ik\underline{\hat{\mathbf{A}}}(\mathbf{r},\omega)\,,
\label{Eperpapp}
\end{equation}
where
\begin{equation}
\bm{\nabla}\cdot\underline{\hat{\mathbf{A}}}(\mathbf{r},\omega)=0\,,
\label{Cgaugepapp}
\end{equation}
and
\begin{equation}
\underline{\hat{\mathbf{E}}}^{\parallel}(\mathbf{r},\omega)=
-\bm{\nabla}\varphi(\mathbf{r},\omega)\,. \label{Eparapp}
\end{equation}
From Eqs.~(\ref{EIapp})--(\ref{Eparapp}), in view of
Eq.~(\ref{elperpar}) and (\ref{Erw}), one obtains the following
equation for the Green tensor components
\begin{equation}
\sum_{\alpha=r,\varphi,z}\!\!\!
\left(\bm{\nabla}\!\times\bm{\nabla}\!\times-\,k^{2}\right)_{\!z\alpha}
\left[\;^{\!\!\perp}G_{\alpha
z}(\mathbf{r},\mathbf{R},\omega)+\;^{\!\!\parallel}G_{\alpha
z}(\mathbf{r},\mathbf{R},\omega)\right]=\delta(\mathbf{r}-\mathbf{R})
\label{Greenperparapp}
\end{equation}
with additional constraints
\begin{equation}
\sum_{\alpha=r,\varphi,z}\!\!\!\nabla_{\alpha}\,^{\!\!\perp}G_{\alpha
z}(\mathbf{r},\mathbf{R},\omega)=0 \label{Gperpgaugeapp}
\end{equation}
and
\begin{equation}
\sum_{\beta,\gamma=r,\varphi,z}\!\!\!
\epsilon_{\alpha\beta\gamma}\nabla_{\beta}\;^{\!\parallel}G_{\gamma
z}(\mathbf{r},\mathbf{R},\omega)=0\,, \label{Gpargaugeapp}
\end{equation}
where $\epsilon_{\alpha\beta\gamma}$ is the totally antisymmetric
unit tensor of rank 3.~Keeping Eqs.~(\ref{Gperpgaugeapp}) and
(\ref{Gpargaugeapp}) in mind, Eq.~(\ref{Greenperparapp}) is
rewritten to give two independent equations for the transverse and
longitudinal Green tensors of the form
\begin{eqnarray}
\left(\Delta+k^{2}\right)\,^{\!\!\perp}G_{zz}(\mathbf{r},\mathbf{R},\omega)\!\!\!
&=&\!\!\!-\delta_{zz}^{\perp}(\mathbf{r}-\mathbf{R}),
\label{Gzzperpapp}\\
k^{2}\;^{\!\parallel}G_{zz}(\mathbf{r},\mathbf{R},\omega)\!\!\!
&=&\!\!\!-\delta_{zz}^{\parallel}(\mathbf{r}-\mathbf{R})
\label{Gzzparapp}
\end{eqnarray}
with the transverse and longitudinal $\delta$-functions given by
Eqs.~(\ref{deltapar}) and (\ref{deltaper}), respectively, and
$^{\!\perp(\parallel)}G_{zz}(\mathbf{r},\mathbf{R},\omega)$
defined by Eq.~(\ref{Gzzperpar}).

Electromagnetic properties of the "atom--nanotube" coupled system
are, according to Eqs.~(\ref{DOS})--(\ref{DOSdef}), determined by
the Green functions
$^{\perp}G_{zz}^{\perp}(\mathbf{r}_{A},\mathbf{r}_{A},\omega)$ and
$^{\parallel}G_{zz}^{\parallel}(\mathbf{r}_{A},\mathbf{r}_{A},\omega)$.~To
derive them, we start with
$^{\perp}G_{zz}(\mathbf{r},\mathbf{r}_{A},\omega)$ for which we
use the differential representation
\begin{equation}
^{\perp}G_{zz}(\mathbf{r},\mathbf{r}_{A},\omega)=\left(
\frac{1}{k^{2}}\,\nabla_{z}\nabla_{z}+1\right)g(\mathbf{r},\mathbf{r}_{A},\omega)
\label{gperapp}
\end{equation}
equivalent to Eq.~(\ref{Gzzperpar}).~Here,
$g(\mathbf{r},\mathbf{r}_{A},\omega)$ is the Green function of the
scalar Helm\-holtz equation which satisfies the radiation
condition at infinity and boundary conditions on the CN
surface.~Indeed, substituting Eq.~(\ref{gperapp}) into
Eq.~(\ref{Gzzperpapp}) and using Eq.~(\ref{Gperpgaugeapp}), one
straightforwardly obtains
\begin{equation}
\left(\Delta+k^{2}\right)g(\mathbf{r},\mathbf{r}_{A},\omega)=
-\delta_{zz}^{\perp}(\mathbf{r}-\mathbf{r}_{A})
\label{gpereqapp}
\end{equation}
-- the scalar Helmholtz equation with a transverse
$\delta$-source. For the longitudinal Green function
$^{\parallel}G_{zz}(\mathbf{r},\mathbf{r}_{A},\omega)$, one
analogously has
\begin{equation}
^{\parallel}G_{zz}(\mathbf{r},\mathbf{r}_{A},\omega)=
-\frac{1}{k^{2}}\,\nabla_{z}\nabla_{z}\,g(\mathbf{r},\mathbf{r}_{A},\omega),
\label{gparapp}
\end{equation}
which upon substituting into Eq.~(\ref{Gzzparapp}) yields
\begin{equation}
\nabla_{z}\nabla_{z}\,g(\mathbf{r},\mathbf{r}_{A},\omega)=
\delta_{zz}^{\parallel}(\mathbf{r}-\mathbf{r}_{A})\,.
\label{gpareqapp}
\end{equation}

Eq.~(\ref{gpereqapp}) has a known solution
\begin{equation}
g_{0}(\mathbf{r},\mathbf{r}_{A},\omega)=\frac{1}{4\pi}\,
\frac{e^{ik|\mathbf{r}-\mathbf{r}_{A}|}}{|\mathbf{r}-\mathbf{r}_{A}|}
\label{g0}
\end{equation}
satisfying the radiation condition at infinity (see,
e.g.,~\cite{Davydov}).~In our case, however, the functions
\begin{equation}
^{\perp}G_{\alpha z}(\mathbf{r},\mathbf{r}_{A},\omega)=\left(
\frac{1}{k^{2}}\,\nabla_{\alpha}\nabla_{z}+\delta_{\alpha
z}\right)g(\mathbf{r},\mathbf{r}_{A},\omega) \label{Gperpalphaz}
\end{equation}
and $g(\mathbf{r},\mathbf{r}_{A},\omega)$, respectively, are
imposed one more set of boundary conditions.~They are the boundary
conditions on the surface of the CN. Using simple relations
\begin{eqnarray}
\underline{E\!}_{\,\alpha}(\mathbf{r},\omega)=ik
^{\,\perp}G_{\alpha z}(\mathbf{r},\mathbf{r}_{A},\omega)
\hskip0.75cm
\label{Ealpha}\\[0.2cm]
\underline{H\!}_{\,\alpha}(\mathbf{r},\omega)=-\frac{i}{k}\!\!\sum_{\beta,\gamma=r,\varphi,z}
\!\!\!\epsilon_{\alpha\beta\gamma}\nabla_{\beta}E_{\gamma}(\mathbf{r},\omega),
\label{Halpha}
\end{eqnarray}
valid at $\mathbf{r}\ne\mathbf{r}_{A}$ for a classical
electromagnetic field under the Coulomb-gauge
condition~\cite{Jackson}, they can be derived from the classical
electromagnetic field boundary conditions of the form
\begin{equation}
\underline{E\!}_{\,\varphi}|_{r=R_{cn}+0}-
\underline{E\!}_{\,\varphi}|_{r=R_{cn}-0}=0\,,
\label{boundaryEphiCN}
\end{equation}
\begin{equation}
\underline{E\!}_{\,z}|_{r=R_{cn}+0}-
\underline{E\!}_{\,z}|_{r=R_{cn}-0}=0\,, \label{boundaryEzCN}
\end{equation}
\begin{equation}
\underline{H\!}_{\,\varphi}|_{r=R_{cn}+0}-
\underline{H\!}_{\,\varphi}|_{r=R_{cn}-0}
=\frac{4\pi}{c}\,\sigma_{zz}(R_{cn},\omega)
{\underline{E\!}_{\,z}}|_{r=R_{cn}}, \label{boundaryHphiCN}
\end{equation}
\begin{equation}
\underline{H\!}_{\,z}|_{r=R_{cn}+0}-
\underline{H\!}_{\,z}|_{r=R_{cn}-0}=0 \label{boundaryHzCN}
\end{equation}
(spatial dispersion neglected) obtained in Ref.~\cite{Slepyan}.

Let $r_{A}\!>\!R_{cn}$ (the atom is outside the CN) to be
specific.~Then, the function
$g(\mathbf{r},\mathbf{r}_{A},\omega)$, being the complete solution
of Eq.~(\ref{gpereqapp}), is represented as a sum of a~particular
solution of the inhomogeneous equation and a general solution of
the homogeneous equation in the form
\begin{equation}
g(\mathbf{r},\mathbf{r}_{A},\omega)=\left\{\begin{array}{ll}
g_{0}(\mathbf{r},\mathbf{r}_{A},\omega)+g^{(+)}(\mathbf{r},\omega)\,,&r>R_{cn}\\[0.1cm]
g^{(-)}(\mathbf{r},\omega)\,,&r<R_{cn}\end{array}\right.
\label{gdef}
\end{equation}
where $g_{0}(\mathbf{r},\mathbf{r}_{A},\omega)$ is the point
radiative atomic source function defined by Eq.~(\ref{g0}) and
$g^{(\pm)}(\mathbf{r},\omega)$ are unknown nonsingular functions
satisfying the homogeneous Helm\-holtz equation and the radiation
conditions at infinity.~We seek them using integral decompositions
over the modified cylindric Bessel functions $I_{p}\,$ and $K_{p}$
as follows~\cite{Jackson}
\begin{equation}
g^{(\pm)}(\mathbf{r},\omega)=\!\!\sum_{p=-\infty}^{\infty}\!\!e^{ip\varphi}\!\!
\int_{C}\left\{\!\!\begin{array}{l}A_{p}(h)\,K_{p}(vr)\\[0.1cm]
B_{p}(h)\,I_{p}(vr)\end{array} \!\!\right\}e^{ihz}dh
\label{gpmexpan}
\end{equation}
and
\begin{equation}
g_{0}(\mathbf{r},\mathbf{r}_{A},\omega)=
\frac{1}{(2\pi)^{2}}\!\!\sum_{p=-\infty}^{\infty}\!\!e^{ip\varphi}\!\!
\int_{C}I_{p}(vr)\,K_{p}(vr_{A})\,e^{ihz}dh\,, \;\;\;r_{A}\geq
r\,, \label{g0expan}
\end{equation}
where $A_{p}(h)$ and $B_{p}(h)$ are unknown functions to be found
from the boundary conditions
(\ref{boundaryEphiCN})--(\ref{boundaryHzCN}) in view of
Eqs.~(\ref{Gperpalphaz}),~(\ref{Ealpha}) and (\ref{Halpha}),
$v\!=\!v(h,\omega)\!=\!\sqrt{h^{2}-k^{2}}$. The integration
contour $C$ runs along the real axis of the complex plane and
envelopes the branch points $\pm k\,$ from below and from above,
respectively. The coordinate system has been fixed as is shown in
Fig.~\ref{fig1}.

The boundary conditions
(\ref{boundaryEphiCN})--(\ref{boundaryHzCN}) with
Eqs.~(\ref{Gperpalphaz}), (\ref{Ealpha}) and (\ref{Halpha}) taken
into account yield the following two independent equations for the
scalar Green function~(\ref{gdef})
\begin{eqnarray}
g_{0}(\mathbf{r},\mathbf{r}_{A},\omega)|_{r=R_{cn}}+
g^{(+)}(\mathbf{r},\omega)|_{r=R_{cn}}
=g^{(-)}(\mathbf{r},\omega)|_{r=R_{cn}},\nonumber\\[0.5cm]
\frac{\partial g^{(+)}(\mathbf{r},\omega)}{\partial
r}|_{r=R_{cn}}-\frac{\partial g^{(-)}(\mathbf{r},\omega)}{\partial
r}|_{r=R_{cn}}\hskip1.5cm\nonumber\\
+\beta(\omega)\!\left(\!\frac{\partial^{2}}{\partial
z^{2}}\!+\!k^{2}\!\right)\!g^{(-)}(\mathbf{r},\omega)|_{r=R_{cn}}\!\!=\!-\frac{\partial
g_{0}(\mathbf{r},\mathbf{r}_{A},\omega)}{\partial
r}|_{r=R_{cn}},\nonumber
\end{eqnarray}
where $\beta(\omega)\!=\!4\pi
i\,\sigma_{zz}(R_{cn},\omega)/\omega$.~Substituting the integral
decompositions (\ref{gpmexpan}) and (\ref{g0expan}) into these
equations, one obtains the set of two simultaneous algebraic
equations for the functions $A_{p}(h)$ and $B_{p}(h)$.~The
function $A_{p}(h)$ we need (we only need the Green function in
the region where the atom is located) is found by solving this set
with the use of basic properties of cylindric Bessel functions
(see, e.g.,~\cite{Abramovitz,Watson}). In so doing, one has
\[
A_{p}(h)=-\frac{R_{cn}\beta(\omega)\,v^{2}I_{p}^{2}(vR_{cn})
K_{p}(vr_{A})}{(2\pi)^{2}[1+\beta(\omega)\,v^{2}R_{cn}
I_{p}(vR_{cn})K_{p}(vR_{cn})]}\,.
\]
These $A_{p}(h)$, being substituted into Eq.~(\ref{gpmexpan}),
yield the function $g^{(+)}(\mathbf{r},\omega)$ sought. The latter
one, in view of Eq.~(\ref{gdef}), results in the scalar
electromagnetic field Green function of the form
\begin{eqnarray}
g(\mathbf{r},\mathbf{r}_{A},\omega)\!\!\!&=&\!\!\!
g_{0}(\mathbf{r},\mathbf{r}_{A},\omega)\nonumber\\[0.3cm]
&-&\!\!\!\!\frac{R_{cn}}{(2\pi)^{2}}\!\!\sum_{p=-\infty}^{\infty}\!\!
e^{ip\varphi}\!\!\int_{C}\!\frac{\beta(\omega)\,v^{2}I_{p}^{2}(vR_{cn})
K_{p}(vr_{A})K_{p}(vr)}{1+\beta(\omega)\,v^{2}R_{cn}
I_{p}(vR_{cn})K_{p}(vR_{cn})}\,e^{ihz}dh, \label{g}
\end{eqnarray}
where $r_{A}\!\ge\!r\!>\!R_{cn}$.~One may show in a similar way
that the function $g(\mathbf{r},\mathbf{r}_{A},\omega)$ for
$r\!\le\!r_{A}\!<\!R_{cn}$ is obtained from Eq.~(\ref{g}) by means
of a simple symbol replacement $I_{p}\!\leftrightarrow\!K_{p}$ in
the numerator of the integrand.

Knowing $g(\mathbf{r},\mathbf{r}_{A},\omega)$, one can easily
calculate from Eq.~(\ref{Gperpalphaz}) the components of the ~
electromagnetic ~ field ~ Green ~ tensor ~ $^{\perp}G_{\alpha
z}(\mathbf{r},\mathbf{r}_{A},\omega)$ ~ and, ~ consequently,
$^{\perp}G_{zz}(\mathbf{r},\mathbf{r}_{A},\omega)$ and
$^{\parallel}G_{zz}(\mathbf{r},\mathbf{r}_{A},\omega)$ defined by
Eqs.~(\ref{gperapp}) and (\ref{gparapp}), respectively, which we
actually need.

\subsubsection*{(b) Local photonic DOS near a single-wall carbon nanotube}

Using Eqs.~(\ref{gperapp}), (\ref{gparapp}), (\ref{Gperpgaugeapp})
and the property
$g(\mathbf{r},\mathbf{r}_{A},\omega)=g(\mathbf{r}_{A},\mathbf{r},\omega)$
which is obvious from Eqs.~(\ref{g0}) and (\ref{g}), it is not
difficult to prove that
\begin{eqnarray}
^{\perp}G_{zz}^{\perp}(\mathbf{r}_{A},\mathbf{r}_{A},\omega)\!\!\!&=&\!\!\!
^{\perp}G_{zz}(\mathbf{r}_{A},\mathbf{r}_{A},\omega)\,,\label{Gzzperperapp}\\
\nonumber\\[-0.4cm]
^{\parallel}G_{zz}^{\parallel}(\mathbf{r}_{A},\mathbf{r}_{A},\omega)\!\!\!&=&\!\!\!
^{\parallel}G_{zz}(\mathbf{r}_{A},\mathbf{r}_{A},\omega)\,,\label{Gzzparparapp}\\
\nonumber\\[-0.4cm]
^{\perp}G_{zz}^{\parallel}(\mathbf{r}_{A},\mathbf{r}_{A},\omega)\!\!\!&=&\!\!\!0\,,\label{Gzzperparapp}
\end{eqnarray}
whereupon, making use of the definition (\ref{DOSdef}) and
Eqs.~(\ref{G0})--(\ref{xipar}), one arrives for $r_{A}>R_{cn}$ at
the $\mathbf{r}_{A}$-dependent transverse and longitudinal local
photonic DOS functions of the form
\begin{eqnarray}
\overline{\xi}^{\,\perp}(\mathbf{r}_{A},\omega)\!\!\!&=&\!\!\!
\frac{3R_{cn}}{2\pi k^{3}}\;\mbox{Im}\!\!\!
\sum_{p=-\infty}^{\infty}\int_{C}\frac{dh\,\beta(\omega)\,v^{4}
I_{p}^{2}(vR_{cn})K_{p}^{2}(vr_{A})}{1+\beta(\omega)\,v^{2}R_{cn}
I_{p}(vR_{cn})K_{p}(vR_{cn})}\,,\label{xiperapp}\\
\overline{\xi}^{\,\parallel}(\mathbf{r}_{A},\omega)\!\!\!&=&\!\!\!\frac{3R_{cn}}{2\pi
k^{3}}\;\mbox{Im}\!\!\!\sum_{p=-\infty}^{\infty}\int_{C}\frac{dh\,\beta(\omega)\,h^{2}v^{2}
I_{p}^{2}(vR_{cn})K_{p}^{2}(vr_{A})}{1+\beta(\omega)\,v^{2}R_{cn}
I_{p}(vR_{cn})K_{p}(vR_{cn})}\,.\label{xiparapp}
\end{eqnarray}
Here, the sign in front of Eq.~(\ref{xiparapp}) has been changed
to be positive.~This reflects the fact that the right-hand sides
of Eqs.~(\ref{gpereqapp}) and (\ref{gpareqapp}) are of opposite
signs.~As a~consequence, partial solutions of the corresponding
homogeneous equations should be taken to have opposite signs as
well.~This yields a correct (positive) sign of the longitudinal
local photonic DOS which, along with the transverse local photonic
DOS, must be a positively defined function.~For $r_{A}<R_{cn}$,
Eqs.~(\ref{xiperapp}) and (\ref{xiparapp}) should be modified by
the replacement $r_{A}\!\leftrightarrow\!R_{cn}$ in the Bessel
function arguments in the numerators of the integrands.~In
dimensionless variables (\ref{omegadimless}), after the
transformation of the integration variable $h\!=\!ky$, these
equations are rewritten to give Eqs.~(\ref{ksiper}) and
(\ref{ksipar}).

\subsubsection*{(c) Proof of Eqs.~(\ref{DOS}) and (\ref{DOSparper})}

Based upon the properties
(\ref{Gzzperperapp})--(\ref{Gzzperparapp}), one can easily prove
Eqs.~(\ref{DOS}) and (\ref{DOSparper}) for the dipole interaction
matrix elements.~The proof of Eq.~(\ref{DOS}) is
straightforward.~In view of Eq.~(\ref{gperppar}), one has
\begin{eqnarray}
\int\!d\mathbf{R}\,|\mbox{g}^{\perp(\parallel)}(\mathbf{r}_{A},\mathbf{R},\omega)|^{2}
\!\!\!&=&\!\!\!\pi\hbar\omega\frac{16\omega_{A}^{2}d_{z}^{\,2}}{c^{4}}\!\int\!d\mathbf{R}\,
\mbox{Re}\,\sigma_{zz}(\mathbf{R},\omega)\nonumber\\
&\times&\!\!\!^{\perp(\parallel)}G_{zz}(\mathbf{r}_{A},\mathbf{R},\omega)
^{\perp(\parallel)}G_{zz}(\mathbf{r}_{A},\mathbf{R},\omega)^{\!\!^{\displaystyle\ast}}.
\label{gperparapp}
\end{eqnarray}
Using further the integral relationship
\begin{equation}
\mbox{Im}\,G_{\alpha\beta}(\mathbf{r},\mathbf{r}^{\prime},\omega)=
\frac{4\pi}{c}\,k\!\int\!d\mathbf{R}\,\mbox{Re}\,\sigma_{zz}(\mathbf{R},\omega)\,
G_{\alpha z}(\mathbf{r},\mathbf{R},\omega)\,G_{\beta
z}(\mathbf{r}^{\prime},\mathbf{R},\omega)^{\!\!^{\displaystyle\ast}}
\label{intrelation}
\end{equation}
[which is nothing but a particular 2D case of a general integral
relationship proven for any 3D electromagnetic field Green tensor
in Ref.~\cite{Welsch}, with Eq.~(\ref{sigmaCN}) taken into
account], one obtains for the right-hand side of
Eq.~(\ref{gperparapp})
\[
\hbar\frac{4\omega_{A}^{2}d_{z}^{\,2}}{c^{2}}\,
\mbox{Im}^{\perp(\parallel)}G^{\perp(\parallel)}_{zz}
(\mathbf{r}_{A},\mathbf{r}_{A},\omega),
\]
whereupon using Eqs.~(\ref{DOSdef}) and (\ref{G0}), one arrives at
Eq.~(\ref{DOS}).

The proof of Eq.~(\ref{DOSparper}) is based upon
Eq.~(\ref{DOS}).~In view of this equation, the right-hand side of
Eq.~(\ref{DOSparper}) takes the form
\begin{eqnarray}
\int\!d\mathbf{R}\,|\mbox{g}^{(\pm)}(\mathbf{r}_{A},\mathbf{R},\omega)|^{2}\!\!\!&=&\!\!\!
\frac{(\hbar\omega_{A})^{2}}{2\pi\omega^{2}}\,\Gamma_{0}(\omega)\!
\left[\xi^{\perp}(\mathbf{r}_{A},\omega)+\!\left(\frac{\omega}{\omega_{A}}\right)^{\!2}
\xi^{\parallel}(\mathbf{r}_{A},\omega)\right]\label{DOSparperapp}\\
&\pm&\!\!\!\frac{\omega}{\omega_{A}}\,2\mbox{Re}\!\int\!d\mathbf{R}\,
\mbox{g}^{\perp}(\mathbf{r}_{A},\mathbf{R},\omega)\,
\mbox{g}^{\parallel}(\mathbf{r}_{A},\mathbf{R},\omega)^{\!\!^{\displaystyle\ast}}\!,\nonumber
\end{eqnarray}
where the last item can be further rewritten in terms of
Eqs.~(\ref{gperppar}), (\ref{Gzzperpar}) and (\ref{intrelation})
to give
\[
\hbar\frac{8\omega\omega_{A}d_{z}^{\,2}}{c^{2}}\,
\mbox{Im}\,^{\perp}G^{\parallel}_{zz}(\mathbf{r}_{A},\mathbf{r}_{A},\omega)\,,
\]
which is zero, according to Eq.~(\ref{Gzzperparapp}).~Thus, one
arrives at Eq.~(\ref{DOSparper}).

\section{Relation of Eq.~(\ref{Evw}) with
the Casimir--Polder formula}\label{Casimir}

In deriving an 'infinitely conducting plane' result (the
Casimir-Polder formula~\cite{Casimir}) from our theory, we will
follow a~general line of the work by Marvin and
Toigo~\cite{Marvin} who did the same within the framework of their
linear response theory.

We start with our weak-coupling-regime
equation~(\ref{Evwweak1}).~First of all one has to simplify the
local DOS functions
$\overline{\xi}^{\,\perp(\parallel)}(\mathbf{r}_{A},x)$ in this
equation by putting $\overline{\sigma}_{zz}\!\rightarrow\infty$.
From Eqs.~(\ref{ksiper}) and (\ref{ksipar}), one has then for
$r_{A}>R_{cn}$
\begin{eqnarray}
\left\{\!\begin{array}{c}
\overline{\xi}^{\,\perp}(\mathbf{r}_{A},x)\\
\overline{\xi}^{\,\parallel}(\mathbf{r}_{A},x)
\end{array}\!\right\}\!\!\!&=&\!\!\!\frac{3}{\pi}\;\mbox{Im}\!\int_{0}^{\infty}\!\!\!
dy\left\{\!\!\begin{array}{c}y^{2}-1-i\varepsilon\\y^{2}\end{array}\!\!\right\}\label{appC1}\\[0.3cm]
&\times&\!\!\!\!\!\sum_{p=-\infty}^{\infty}\!\!\!\frac{K_{p}^{2}[\sqrt{y^{2}-1-i\varepsilon}\,u(r_{A})x]
I_{p}[\sqrt{y^{2}-1-i\varepsilon}\,u(R_{cn})x]}{K_{p}[x\sqrt{y^{2}-1-i\varepsilon}\,u(R_{cn})x]}\,,
\nonumber
\end{eqnarray}
where $\varepsilon$ is an infinitesimal positive constant which is
necessary to correctly envelope the branch points of the integrand
in integrating over $y$.

The next step is taking a large radius limit in Eq.~(\ref{appC1}).
This can be done by using the relationship
\[
\lim_{\begin{array}{c}\mbox{\scriptsize$a\!\rightarrow\!\infty$}\\[-0.15cm]
\mbox{\scriptsize$b\!\rightarrow\!\infty$}\\[-0.15cm]
\mbox{\scriptsize$(a\!-\!b=const)$}\end{array}}\!\!\!\!\!\!\!
\sum_{n=-\infty}^{\infty}\!\!\!\frac{K_{n}^{2}(a)I_{n}(b)}{K_{n}(b)}=K_{0}[2(a-b)]
\]
proved in Ref.~\cite{Marvin}.~One has
\begin{equation}
\left\{\!\begin{array}{c}
\overline{\xi}^{\,\perp}(\mathbf{r}_{A},x)\\
\overline{\xi}^{\,\parallel}(\mathbf{r}_{A},x)\end{array}\!\right\}
=\frac{3}{\pi}\;\mbox{Im}\!\int_{0}^{\infty}\!\!\!dy\left\{\!\!
\begin{array}{c}y^{2}-1-i\varepsilon\\y^{2}\end{array}\!\!\right\}
K_{0}[2\mu x\sqrt{y^{2}-1-i\varepsilon}\;]\,, \label{appC2}
\end{equation}
where $\mu\!=\!2\gamma_{0}l/\hbar c$ with $l\!=\!r_{A}-R_{cn}$
being the atom-surface distance.~The integration variable $y$ in
this equation contains implicit frequency dependence.~Indeed,
comparing dimensionless equations (\ref{ksiper}), (\ref{ksipar})
with (\ref{xiperapp}), (\ref{xiparapp}), one can see that
$y=h/k=h(c/\omega)=h(c\hbar/2\gamma_{0}x)$. For the following it
makes sense to extract this frequency dependence by the
substitution $y\!=\!\lambda/x$ with the dimensionless
$\lambda=h(c\hbar/2\gamma_{0})$.~The result takes the form
\begin{equation}
\overline{\xi}^{\,\perp(\parallel)}(\mathbf{r}_{A},x)=
\frac{3}{2i\pi x}\left\{f^{\perp(\parallel)}(\mathbf{r}_{A},x)-
\!f^{\perp(\parallel)}(\mathbf{r}_{A},x)^{\!\!^{\displaystyle\ast}}\right\}
\label{appC3}
\end{equation}
with
\begin{equation}
\left\{\!\begin{array}{c}
f^{\perp}(\mathbf{r}_{A},x)\\
f^{\parallel}(\mathbf{r}_{A},x)\end{array}\!\right\}
=\int_{0}^{\infty}\!\!\!d\lambda\left\{\!\!
\begin{array}{c}(\lambda/x)^{2}-1-i\varepsilon\\
(\lambda/x)^{2}\end{array}\!\!\right\}K_{0}[2\mu
x\sqrt{(\lambda/x)^{2}- 1-i\varepsilon}\;]\,.\label{appC4}
\end{equation}

Substituting Eq.~(\ref{appC3}) into Eq.~(\ref{Evwweak1}), one has
\begin{eqnarray}
\varepsilon_{vw}(\mathbf{r}_{A})\!\!\!&=&\!\!\!\frac{3x_{A}}{4i\pi^{2}}\left\{
\int_{0}^{\infty}\!\!\!\frac{dx\,\tilde{\Gamma}_{0}(x)}{(x_{A}\!+x)x^{2}}\!
\left[f^{\perp}(\mathbf{r}_{A},x)\!-\!\frac{x}{x_{A}}f^{\parallel}(\mathbf{r}_{A},x)\right]
\right.\label{appC5}\\
&-&\!\!\!\left.\int_{0}^{-\infty}\!\!\!\frac{dx\,\tilde{\Gamma}_{0}(x)}{(x_{A}\!-x)x^{2}}\!
\left[f^{\perp}(\mathbf{r}_{A},-x)^{\!\!^{\displaystyle\ast}}\!+\!
\frac{x}{x_{A}}f^{\parallel}(\mathbf{r}_{A},-x)^{\!\!^{\displaystyle\ast}}\right]
\!\right\}.\nonumber
\end{eqnarray}
Here, in the second item, the change of the integration variable
from $x$ to $-x$ has been made and it has been taken into account
that $\tilde{\Gamma}_{0}(-x)\!=\!-\tilde{\Gamma}_{0}(x)$.~Noting
further that both integrands in Eq.~(\ref{appC5}) do not have
poles in the upper complex half plane, one may rotate the
integration path of the first and the second integral by $\pi/2$
and by $-\pi/2$, respectively.~Then, both integrals become the
integrals over the positive imaginary axis of the complex plane
and Eq.~(\ref{appC5}) takes the form
\begin{eqnarray}
\varepsilon_{vw}(\mathbf{r}_{A})\!\!\!&=&\!\!\!\frac{3x_{A}}{4i\pi^{2}}\left\{
\int_{0}^{i\infty}\!\!\!\frac{dz\,\tilde{\Gamma}_{0}(z)}{(x_{A}\!+z)z^{2}}\!
\left[f^{\perp}(\mathbf{r}_{A},z)\!-\!\frac{z}{x_{A}}f^{\parallel}(\mathbf{r}_{A},z)\right]
\right.\mbox{\hskip0.5cm}\label{appC6}\\
&-&\!\!\!\left.\int_{0}^{i\infty}\!\!\!\frac{dz\,\tilde{\Gamma}_{0}(z)}{(x_{A}\!-z)z^{2}}\!
\left[f^{\perp}(\mathbf{r}_{A},-z)^{\!\!^{\displaystyle\ast}}\!+\!
\frac{z}{x_{A}}f^{\parallel}(\mathbf{r}_{A},-z)^{\!\!^{\displaystyle\ast}}\right]
\!\right\}.\nonumber
\end{eqnarray}
Making further the change of the integration variable from $z$ to
$iu$ with real non-negative $u$, one arrives at
\begin{eqnarray}
\varepsilon_{vw}(\mathbf{r}_{A})\!\!\!&=&\!\!\!\frac{3x_{A}}{4\pi^{2}}\left\{
-\!\int_{0}^{\infty}\!\!\!\frac{du\,\tilde{\Gamma}_{0}(iu)}{(x_{A}\!+iu)u^{2}}\!
\left[f^{\perp}(\mathbf{r}_{A},iu)\!-\!\frac{iu}{x_{A}}f^{\parallel}(\mathbf{r}_{A},iu)\right]
\right.\mbox{\hskip0.5cm}\label{appC7}\\
&+&\!\!\!\left.\int_{0}^{\infty}\!\!\!\frac{du\,\tilde{\Gamma}_{0}(iu)}{(x_{A}\!-iu)u^{2}}\!
\left[f^{\perp}(\mathbf{r}_{A},-iu)^{\!\!^{\displaystyle\ast}}\!+\!
\frac{iu}{x_{A}}f^{\parallel}(\mathbf{r}_{A},-iu)^{\!\!^{\displaystyle\ast}}\right]
\!\right\}.\nonumber
\end{eqnarray}
Here, the functions $f^{\perp(\parallel)}(\mathbf{r}_{A},iu)$ have
the following explicit form
\begin{eqnarray}
\left\{\!\!\begin{array}{c}
f^{\perp}(\mathbf{r}_{A},iu)\\
f^{\parallel}(\mathbf{r}_{A},iu)\end{array}\!\!\right\}
=-u\!\!\int_{1}^{\infty}\!\!\!\!\!d\eta\left\{\!\!
\begin{array}{c}\eta^{3}\!/\sqrt{\eta^{2}-1}\\
\eta\sqrt{\eta^{2}-1}\end{array}\!\right\}\!K_{0}(2\mu u\eta)
\label{appC8}
\end{eqnarray}
obtained from Eq.~(\ref{appC4}) after the change of the
integration variable by the subsequent substitutions $x=iu$ and
$\eta^{2}\!=\!\lambda^{2}/u^{2}+1$.~The functions
$f^{\perp(\parallel)}(-iu)^{\ast}\! =\!f^{\perp(\parallel)}(iu)$
as a~consequence~of a general property
$K_{\nu}(z)^{\ast}\!=\!K_{\nu}(z^{\ast})$~\cite{Abramovitz}.
Substituting Eq.~(\ref{appC8}) into Eq.~(\ref{appC7}) and making
allowance for the fact that
$\tilde{\Gamma}_{0}(iu)\!=\!-i\tilde{\Gamma}_{0}(u)$, one further
has
\begin{equation}
\varepsilon_{vw}(\mathbf{r}_{A})=-\frac{3x_{A}}{2\pi^{2}}\!
\int_{0}^{\infty}\!\frac{du\,\tilde{\Gamma}_{0}(u)}{x_{A}^{2}\!+u^{2}}
\int_{1}^{\infty}\!\!\!d\eta\,\eta\!\left(\!\sqrt{\eta^{2}-1}
+\frac{\eta^{2}}{\sqrt{\eta^{2}-1}}\right)\!K_{0}(2\mu u\eta)\,,
\label{appC9}
\end{equation}
which upon the substitution $\eta^{2}=\rho$ becomes
\begin{equation}
\varepsilon_{vw}(\mathbf{r}_{A})=-\frac{3x_{A}}{4\pi^{2}}\!
\int_{0}^{\infty}\!\frac{du\,\tilde{\Gamma}_{0}(u)}{x_{A}^{2}\!+u^{2}}
\int_{1}^{\infty}\!\!\!d\rho\left(\!\sqrt{\rho-1}
+\frac{\rho}{\sqrt{\rho-1}}\right)\!K_{0}(2\mu u\sqrt{\rho}\,)\,.
\label{appC10}
\end{equation}

In Eq.~(\ref{appC10}), the first integral over $\rho$ is taken
exactly by means of the relationship~\cite{Ryzhik}
\[
\int_{1}^{\infty}\!\!\!\!\!d\rho\sqrt{\rho-1}\,K_{0}(a\sqrt{\rho}\,)\!=\!
\sqrt{\frac{2\pi}{a^{3}}}\,K_{3/2}(a)
\longrightarrow_{^{^{^{^{^{^{^{\mbox{\hskip-0.75cm}}}}}}}}a\rightarrow\infty}
\frac{\pi}{a^{3}}(1+a)e^{-a}
\]
and the second one -- by making a large argument series expansion
of the integrand and its subsequent termwise integration.~This,
upon returning back to dimensional variables by means of
Eqs.~(\ref{energydimless}) and using the definition (see,
e.g.,~\cite{Davydov,Agarwal})
\[
\alpha_{zz}(iu)=\frac{2}{\hbar}\,\frac{d_{z}^{\,2}\,\omega_{A}}{\omega_{A}^{2}-(iu)^{2}}
\]
for the polarizability tensor $zz$-component of the two-level
system, finally yields
\[
E_{vw}(\mathbf{r}_{A})\approx-\frac{\hbar}{8\pi l^{3}}
\int_{0}^{\infty}\!\!\!du\,\alpha_{zz}(iu)\!
\left[1+2\frac{u}{c}\,l+2\left(\frac{u}{c}\right)^{2}l^{2}\right]e^{-2(u/c)l}.
\]
This is the half of the well-known Casimir-Polder result for the
vdW energy of an atom near an infinitely conducting plane (see
Refs.~\cite{Casimir,Marvin}), in agreement with the fact that our
model of the atom-electromagnetic-field interactions in the
presence of a nanotube only takes the longitudinal (along the
nanotube axis, or, equivalently, parallel to the plane in the
large nanotube radius limit) atomic dipole orientation into
account.~Another half of the Casimir-Polder vdW energy would come
from the transverse (perpendicular to the plane) atomic dipole
orientation which we have neglected.


\begin{thebibliography}{99}
\bibitem{Dresselhaus}Dresselhaus M.S., Dresselhaus G., and Eklund P.C.~~
\emph{Science of Fullerens and Carbon Nanotubes}, Academic Press,
New York, 1996.

\bibitem{Dai}Dai H.~~ \emph{Surf. Sci.} \textbf{500}, 218 (2002).

\bibitem{Baughman}Baughman R.H., Zakhidov A.A., and de Heer W.A.~~
\emph{Science} \textbf{297}, 787 (2002).

\bibitem{Duclaux}Duclaux L.~~ \emph{Carbon} \textbf{40}, 1751 (2002).

\bibitem{Jeong}Jeong G.-H., Farajian A.A., Hatakeyama R., Hirata T., Yaguchi T.,
Tohji K., Mizuseki H., and Kawazoe Y.~~ \emph{Phys. Rev. B}
\textbf{68}, 075410 (2003).

\bibitem{Shimoda}Shimoda H., Gao B., Tang X.P., Kleinhammes A., Fleming L.,
Wu Y., and Zhou O.~~ \emph{Phys. Rev. Lett.} \textbf{88}, 015502
(2002).

\bibitem{Calbi}Calbi M.M., Cole M.W., Gatica S.M., Bojan M.J., and Stan G.~~
\emph{Rev. Mod. Phys.} \textbf{73}, 857 (2001).

\bibitem{Purcell}Purcell E.M.~~ \emph{Phys. Rev.} \textbf{69}, 681 (1946).

\bibitem{Weisbuch}Rarity J.D. and Weisbuch C.~~ \emph{Microcavities
and Photonic Bandgaps: Physics and Applications}, NATO ASI Series,
Vol.~E324, Kluwer, Dordrecht, 1996.

\bibitem{Pelton}Pelton M. and Yamamoto Y.~~ \emph{Phys. Rev. A} \textbf{59}, 2418 (1999).

\bibitem{Vuchkovich}Vu$\check{\mbox{c}}$kovi\'{c} J., Fattal D., Santori C.,
Solomon G.S., and Yamamoto Y.~~ \emph{Appl. Phys. Lett.}
\textbf{82}, 3596 (2003).

\bibitem{Dung}Dung H.T., Kn\"{o}ll L., and Welsch D.-G.~~ \emph{Phys. Rev.
A} \textbf{62}, 053804 (2000); \emph{ibid.} \textbf{64}, 013804
(2001) [and refs. therein].

\bibitem{Welsch}Kn\"{o}ll L., Scheel S., and Welsch D.-G.~~ in:
\emph{Coherence and Statistics of Photons and Atoms}, edited by
Pe$\check{\mbox{r}}$ina J., Wiley, New York, 2001 [see also
\emph{quant-ph}/0006121 (26 Jun 2003)].

\bibitem{Kimble}Buck J.R. and Kimble H.J.~~ \emph{Phys. Rev. A} \textbf{67},
033806 (2003).

\bibitem{Klimov}Klimov V.V. and Ducloy M.~~ \emph{Phys. Rev. A} \textbf{69}, 013812 (2004).

\bibitem{John}Florescu M. and John S.~~ \emph{Phys. Rev. A} \textbf{64}, 033801 (2001).

\bibitem{Sugawara}Sugawara M.~~ \emph{Phys. Rev. B} \textbf{51}, 10743 (1995).

\bibitem{Schniepp}Schniepp H. and Sandoghdar V.~~ \emph{Phys. Rev. Lett.}
\textbf{89}, 257403 (2002).

\bibitem{Reithmaier}Reithmaier J.P., Sek G., L\"{o}ffler A., Hoffman C.,
Kuhn S., Reitzenstein S., Keldysh L.V., Kulakovskii V.D., Reinecke
T.L., and Forchel A.~~ \emph{Nature}, \textbf{432}, 197 (2004).

\bibitem{Gayral}Gayral B., G\'{e}rard J.-M., Sermage B., Lema\^{i}tre A., and Dupuis
C.~~ \emph{Appl. Phys. Lett.} \textbf{78}, 2828 (2001).

\bibitem{Peter}Peter E., Senellart P., Martrou D., Lema$\hat{\mbox{i}}$tre
A., Hours J., G\'{e}rard J.M., and Bloch J.~~
\emph{quant-ph}/0411076 (3 Dec 2004).

\bibitem{Petrov}Petrov E.P., Bogomolov V.N., Kalosha I.I., and Gaponenko
S.V.~~ \emph{Phys. Rev. Lett.} \textbf{81}, 77 (1998).

\bibitem{Yoshie}Yoshie T., Scherer A., Hendrikson J., Khitrova G., Gibbs H.M.,
Rupper G., Ell~C., Shchekin O.B., and Deppe D.G.~~ \emph{Nature},
\textbf{432}, 200 (2004).

\bibitem{Finley}Kress A., Hofbauer F., Reinelt N., Krener H.J., Meyer R.,
B\"{o}hm G., and Finley~J.J.~~ \emph{quant-ph}/0501013 (4 Jan
2005).

\bibitem{Eberly}Allen L. and Eberly J.H.~~ \emph{Optical Resonance and
Two-Level Atoms}, Wiley, New York, 1975.

\bibitem{Bondarev02}Bondarev I.V., Slepyan G.Ya., and Maksimenko S.A.~~
\emph{Phys. Rev. Lett.} \textbf{89}, 115504 (2002).

\bibitem{Bondarev04PLA}Bondarev I.V. and Lambin Ph.~~ \emph{Phys. Lett. A}
\textbf{328}, 235 (2004).

\bibitem{Bondarev04PRB}Bondarev I.V. and Lambin Ph.~~ \emph{Phys. Rev. B}
\textbf{70}, 035407 (2004).

\bibitem{Buhmann04JOB}Buhmann S.Y., Dung H.T., and Welsch D.-G.~~
\emph{J. Opt. B: Quantum Semiclass. Opt.} \textbf{6}, S127 (2004).

\bibitem{Buhmann04pre}Buhmann S.Y., Dung H.T., Kn\"{o}ll L., and Welsch D.-G.~~
\emph{quant-ph}/0403128 (17 Mar 2004).

\bibitem{Zhao2002}Zhao J., Buldum A., Han J., and Lu J.P.~~ \emph{Nanotechnology}
\textbf{13} 195 (2002).

\bibitem{Zhang2003}Zhang X.R., Cao D.P., and Chen J.F.~~ \emph{J. Phys. Chem. B} \textbf{107},
4942 (2003).

\bibitem{Han2004}Han S.S. and Lee H.M.~~ \emph{Carbon} \textbf{42}, 2169 (2004).

\bibitem{Dag2003}Dag S., Gulseren O., Yildirim T., and Ciraci S.~~ \emph{Phys. Rev. B}
\textbf{67}, 165424 (2003).

\bibitem{Girifalco}Girifalco L.A. and Hodak M.~~ \emph{Phys. Rev. B} \textbf{65}, 125404
(2002).

\bibitem{London}London F.~~ \emph{Zs. f. Physik} \textbf{60}, 491 (1930).

\bibitem{Williams2000}Williams K.A. and Eklund P.C.~~ \emph{Chem. Phys. Lett.} \textbf{320}, 352
(2000).

\bibitem{Ulbricht2002}Ulbricht H., Moos G., and Hertel T.~~ \emph{Phys. Rev. B} \textbf{66},
075404 (2002).

\bibitem{Slepyan}Slepyan G.Ya., Maksimenko S.A., Lakhtakia A., Yevtushenko O.,
and Gusa\-kov~A.V.~~ \emph{Phys. Rev. B} \textbf{60}, 17136
(1999).

\bibitem{Benedict}Benedict L.X., Louie S.G., and Cohen M.L.~~ \emph{Phys. Rev. B}
\textbf{52}, 8541 (1995).

\bibitem{Tasaki}Tasaki S., Maekawa K., and Yamabe T.~~ \emph{Phys. Rev. B}
\textbf{57}, 9301 (1998).

\bibitem{Jorio}Jorio A., Souza Filho A.G., Brar V.W., Swan A.K., \"{U}nl\"{u} M.S.,
Goldberg B.B., Righi A., Hafner J.H., Lieber C.M., Saito R.,
Dresselhaus G., and Dresselhaus~M.S.~~ \emph{Phys. Rev. B}
\textbf{65}, 121402(R) (2002).

\bibitem{Li}Li Z.M., Tang Z.K., Liu H.J., Wang N., Chan C.T., Saito R.,
Okada S., Li~G.D., Chen J.S., Nagasawa N., and Tsuda S.~~
\emph{Phys. Rev. Lett.} \textbf{87}, 127401 (2001).

\bibitem{Marinop}Marinopoulos A.G., Reining L., Rubio A., and Vast N.~~
\emph{Phys. Rev. Lett.} \textbf{91}, 046402 (2003).

\bibitem{Weisbuch1}Weisbuch C., Nishioka M., Ishikawa A., and Arakawa Y.~~
\emph{Phys. Rev. Lett.} \textbf{69}, 3314 (1992).

\bibitem{Sorba}Dini D., K\"{o}hler R., Tredicucci A., Biasiol G., and Sorba L.~~
\emph{Phys. Rev. Lett.} \textbf{90}, 116401 (2003).

\bibitem{Bondarev04SSC}Bondarev I.V. and Lambin Ph.~~ \emph{Solid State Commin. A}
\textbf{132}, 203 (2004).

\bibitem{Bondarev05PRB}Bondarev I.V. and Lambin Ph.~~ \emph{cond-mat}/0410216 (8 Oct 2004).

\bibitem{Davydov}Davydov A.S.~~ \emph{Quantum Mechanics}, Pergamon, New York, 1976.

\bibitem{Vogel}Vogel W., Welsch D.-G., and Wallentowitz S.~~ \emph{Quantum Optics:
an Introduction}, Wiley-VCH, New York, 2001.

\bibitem{Jackson}Jackson J.D.~~ \emph{Classical Electrodynamics}, Wiley, New York,
1975.

\bibitem{Abrikosov}Abrikosov A.A., Gorkov L.P., and Dzyaloshinski I.E.~~
\emph{Methods of Quantum Field Theory in Statistical Physics},
Dover, New York, 1975.

\bibitem{Barnett}Barnett S.M., Huttner B., and Loudon R.~~ \emph{Phys. Rev. Lett.}
68, 3698 (1992).

\bibitem{Agarwal75}Agarwal G.S.~~ \emph{Phys. Rev. A} \textbf{12}, 1475 (1975).

\bibitem{Wallace}Wallace P.R.~~ \emph{Phys. Rev.} \textbf{71}, 622 (1947).

\bibitem{Abramovitz}Abramovitz M. and Stegun I.A. (Eds) \emph{Handbook of Mathematical
Functions}, Dover, New York, 1972.

\bibitem{Lambin}Henrard L. and Lambin Ph.~~ \emph{J. Phys. B} \textbf{29}, 5127 (1996).

\bibitem{Heitler}Heitler W.~~ \emph{The Quantum Theory of
Radiation}, Oxford, London, 1954.

\bibitem{Watson}Watson, G.N.~~ \emph{Theory of Bessel Functions},
Cambridge University Press, Cambridge, 1922.

\bibitem{Casimir}Casimir H.B.G. and Polder D.~~ \emph{Phys. Rev.} \textbf{73}, 360 (1948).

\bibitem{Lide}Lide D.R. (Editor-in-Chief)~~ \emph{Handbook of Chemistry
and Physics}, CRC Press, New York, 1999.

\bibitem{Masenelli}Masenelli-Varlot K., McRae E., and Dupont-Pavlovsky N.~~
\emph{Appl. Surf. Sci.} \textbf{196}, 209 (2002).

\bibitem{Haroche1}Haroche S. and Kleppner D.~~ \emph{Phys. Today} \textbf{42}, No. 1, 24
(1989).

\bibitem{Yamamoto}Vu$\check{\mbox{c}}$kovi\'{c} J. and Yamamoto Y.~~ \emph{Appl. Phys. Lett.}
\textbf{82}, 2374 (2003).

\bibitem{Noda}Asano T. and Noda S.~~ \emph{Nature} \textbf{429},
6988 (2004).

\bibitem{Raimond}Raimond J.M., Brune M., and Haroche S.~~ \emph{Rev. Mod.
Phys.} \textbf{73}, 565 (2001).

\bibitem{Abstreiter}Zrenner A., Beham E., Stufler S., Findeis F., Bichler M.,
and Abstreiter G.~~\linebreak \emph{Nature} \textbf{418}, 612
(2002).

\bibitem{Steel}Li X., Wu Y., Steel D., Gammon D., Stievater T.H., Katzer D.S.,
Park D., Piermarocchi C., and Sham L.J.~~ \emph{Science}
\textbf{301}, 809 (2003).

\bibitem{Bondarev03}Bondarev I.V., Maksimenko S.A., Slepyan G.Ya., Krestnikov I.L.,
and Hoffmann A.~~ \emph{Phys. Rev. B} \textbf{68}, 073310 (2003).

\bibitem{Scheel}Dung H.T., Scheel S., Welsch D.-G., and Knoll L.~~ \emph{J. Opt. B:
Quantum Semiclass. Opt.} \textbf{4}, S169 (2002).

\bibitem{Agarwal}Agarwal G.S. and Dutta~Gupta S.~~ \emph{Phys. Rev. A} \textbf{57},
667 (1998).

\bibitem{Dung02}Dung H.T., Kn\"{o}ll L., and Welsch D.-G.~~ \emph{Phys. Rev. A} \textbf{66},
063810 (2002).

\bibitem{Dalton}Dalton B.J. and Garraway B.M.~~ \emph{Phys. Rev. A} \textbf{68},
033809 (2003).

\bibitem{Marvin}Marvin A.M. and Toigo F.~~ \emph{Phys. Rev. A} \textbf{25}, 782 (1982).

\bibitem{Ryzhik}Gradshteyn I.S. and Ryzhik I.M.~~ \emph{Tables of Integrals, Series
and Products}, Academic, New York, 1965.
\end{thebibliography}
\end{document}